\documentclass[12pt,preprint]{aastex}
\usepackage{graphicx}
 
\begin{document}

\title{X-rays in the Orion Nebula Cluster: Constraints on the \\
       origins of magnetic activity in pre-main sequence stars}
       
\author{Eric D.\ Feigelson\altaffilmark{1}, James A.\ Gaffney
III\altaffilmark{1}, Gordon Garmire\altaffilmark{1}, Lynne A.\
Hillenbrand\altaffilmark{2}} 
\and 
\author{Leisa Townsley\altaffilmark{1}}

\altaffiltext{1}{Department of Astronomy \& Astrophysics,
525 Davey Laboratory, Pennsylvania State University, University Park
PA 16802}
\altaffiltext{2}{Department of Astronomy, MS 105-24, California
Institute of Technology, Pasadena CA 91125}

\slugcomment{To appear in the Astrophysical Journal}

\begin{abstract}

A recent observation of the Orion Nebula Cluster with the ACIS
instrument on board the $Chandra$ X-ray Observatory detected 1075
sources (Feigelson et al.\ 2002), 525 of which are pre-main sequence
(PMS) stars with measured bulk properties (bolometric luminosities,
masses, ages and disk indicators). Nearly half of these stars have
photometrically measured rotational periods.  This provides a uniquely
large and well-defined sample to study the dependence of magnetic
activity on bulk properties for stars descending the Hayashi tracks.

The following results are obtained:  (1)  X-ray luminosities $L_t$ in
the $0.5-8$ keV band are strongly correlated with bolometric luminosity
with average ratio $\log L_t/L_{bol} = -3.8$ for stars with masses $0.7
< M < 2$ M$_\odot$, an order of magnitude below the main sequence
saturation level;  (2) the X-ray emission drops rapidly below this
level in some or all stars with $2 < M < 3$ M$_\odot$; (3) the presence
or absence of infrared circumstellar disks has no apparent relation to
X-ray levels; and (4) X-ray luminosities exhibit a slight rise as
rotational periods increase from 0.4 to 20 days.  This last finding
stands in dramatic contrast to the strong anticorrelation between
X-rays and period seen in main sequence stars.

The absence of a strong X-ray/rotation relationship in PMS stars, and
particularly the high X-ray values seen in some very slowly rotating
stars, is a clear indication that the mechanisms of magnetic field
generation differ from those operating in main sequence stars.  The
most promising possibility is a turbulent dynamo distributed throughout
the deep convection zone, but other models such as $\alpha-\Omega$
dynamo with `supersaturation' or relic core fields are not immediately
excluded.  The drop in magnetic activity in intermediate-mass stars may
reflect the presence of a significant radiative core.  The evidence
does not support X-ray production in large-scale star-disk magnetic
fields.

\end{abstract}


\newpage

\section{Introduction  \label{intro.sec}}

The astrophysical origin of the surface magnetic activity of solar-type
main sequence stars has been established with some confidence
\citep{Schrijver00}. Magnetic fields are generated by differential
rotation at the interface (tachocline) between the radiative and
convective zones and rise through the convection zone to the surface
where they produce active regions, violent flares, coronal heating and
other effects.  Oscillations in this dynamo account for the 22-year
solar cycle. In other main sequence stars, the principal evidence for
such a dynamo is the ubiquitous relationship between magnetic activity
indicators and surface rotation:  more rapidly rotating stars exhibit
higher levels of activity until, for some indicators, a saturation of
the process is reached.

It is not clear, however, whether this model applies to late-type stars
with substantially different internal structure from the Sun's such as
PMS T Tauri stars, post-main sequence giants, and low-mass M dwarfs.
Such stars may have tachoclines buried deep in the interior or may be
fully convective without any tachocline.  Yet both T Tauri stars and dM
main sequence stars exhibit large active regions and strong flaring
indicating that magnetic field generation is operative.  Various
suggestions have been made to account for this, such as a turbulent
dynamo process distributed throughout the convective zone, but with
little empirical support.  Notably, an activity-rotation relationship
is sometimes but not always evident in these stars.  But the samples for
study have generally been small and the empirical results often
discrepant.

It has proved difficult to study the origins of magnetic activity
in PMS stars using traditional optical and ultraviolet indicators due
to obscuration and confusion arising from gas infall and ejecta.  
Elevated levels of X-ray emisison, in contrast, are ubiquitous in PMS
stars and are relatively unaffected by such problems \citep[see review 
by][]{Feigelson99}.  However, despite considerable effort with the 
$Einstein$ and $ROSAT$ missions, the observational basis for 
understanding the elevated levels of PMS magnetic activity is still
murky.  Some studies show an X-ray/rotation correlation while
others do not, and other confounding  correlations with bulk properties
are present (\S 2.1.3).  The theoretical issues are also more complex
than with main seqence stars (\S 2.2).  

The $Chandra$ ACIS study of the Orion Nebula Cluster (ONC), which
illuminates the M~42 blister HII region on a near edge of the Orion
molecular cloud, provides a unique opportunity to study these issues.
Here, a single image reveals $\sim 1000$ X-ray emitting PMS stars that
span the entire initial mass function and a good portion of the PMS
evolutionary tracks.  The ONC has been the subject of intense optical
and infrared study so that nearly a thousand of its members have been
placed on the Hertzsprung-Russell (HR) diagram and over four hundred
have photometrically measured rotation periods.  Together, the
$Chandra$ and optical results give a great increase in sample size for
study of the origins of PMS magnetic activity compared to previous
efforts. 

We find no evidence for the X-ray/rotation correlation strongly present
in main sequence stars among ONC stars.  Various other effects are
found which may constrain alternative explanations for PMS magnetic
activity.  The most promising interpretation, in our view, is that the
magnetic fields are produced by a distributed dynamo within the deep
convective zone.  Further development of theoretical models is needed
in light of our observational results.

The paper begins with a review of the complex empirical and theoretical
issues concerning magnetic activity and rotation in late-type stars (\S
\ref{xray_dyn.sec}). The Chandra ONC dataset is presented (\S
\ref{data.sec}) and the effects of various stellar properties on the
X-ray emission are explored (\S \ref{depend.sec}).  X-ray/rotation
relations are presented in \S \ref{rotation.sec} followed by discussion
(\S \ref{disc.sec}) and conclusions (\S \ref{concl.sec}). This is the
fourth paper in a series on the Chandra observation of the ONC using
the ACIS-I detector:  \citet{Garmire00} introduced the field and
discussed stars in the BN/KL region; \citet[][henceforth
F02a]{Feigelson02a} give comprehensive tables of the 1075 sources and
discuss X-ray emission as a function of mass; and \citet[][henceforth
F02b]{Feigelson02b} discuss flaring in pre-main sequence analogs of the
early Sun and their implications for the early solar system.

\section{Stellar X-rays and dynamos \label{xray_dyn.sec}}

We review here past observational (\S \ref{xray_rev.sec}) and
theoretical studies (\S \ref{theor_rev.sec}) studies which provide the
foundation for the present study.  We find that the situation for main
sequence F-K stars is reasonably clear: rotation appears to be the
principal observable correlate to X-ray luminosity and, through the
Rossby number, rotation can be linked to an $\alpha-\Omega$-type dynamo
that successfully explains many features of solar and stellar
activity.   The Rossby number $Ro = P/\tau_c$, the ratio of the
rotational period $P$ to the convective overturn time $\tau_c$ near the
base of the stellar convection zone, is a measure of the growth rate
of the field in many dynamo theories.  Rossby numbers account
for mass-dependent structural differences in stellar interiors and are
quite stable to reasonable variations in assumptions concerning the
physics of the convection zone \citep{Montesinos01}.

The situation is more confused for giants and dM stars where only weak
activity/rotation relationships are seen.  It is not clear whether
magnetic fields in the these stars with deep convective zones arise
from a modified $\alpha-\Omega$ dynamo or a distributed turbulent
dynamo.  For PMS stars, the interpretation is even more uncertain:
several dynamo concepts compete with the possibility that the magnetic
fields are inherited from the gravitational collapse or arise from
star-disk interactions.

\subsection{Relationship between stellar X-rays and rotation
\label{xray_rev.sec}}

\subsubsection{Solar-type main sequence stars \label{main_seq.sec}}

The surface magnetic activity of solar-type stars arises from the
emergence and reconnection of fields generated in the stellar interior
\citep[see][for a thorough review]{Schrijver00}.  In the X-ray band,
this consists of a slowly varying soft X-ray corona and hard emission
from violent magnetic reconnection during flares.  The first X-ray
surveys of late-type stars with  the $Einstein$ $Observatory$ revealed
a strong X-ray/rotation correlation of the form $L_s =
10^{27}~(v~sini)^2$ erg s$^{-1}$ where $L_s$ is measured in the soft
$0.5-2.5$ keV band and $v~(sini)$ is the projected rotation speed in km
s$^{-1}$ \citep{Pallavicini81}.  The X-ray/rotation connection for main
sequence stars was repeatedly confirmed in many $Einstein$ and $ROSAT$
studies of both field and open cluster stars. 

\begin{figure}
\centering
  \includegraphics[height=0.45\textheight]{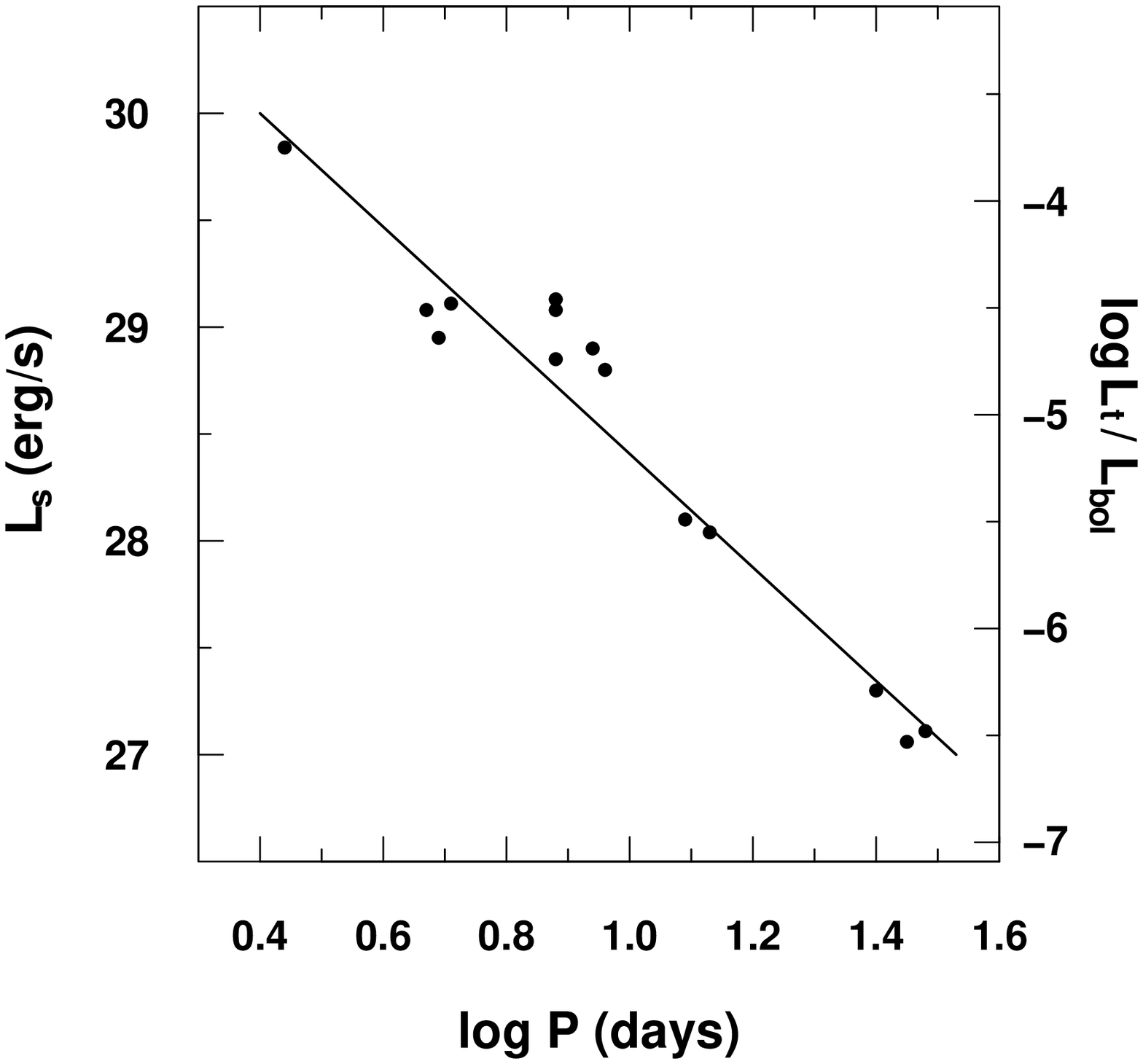}
\caption{Relationships between soft X-ray emission and rotation in main
sequence stars from $ROSAT$ studies:  (a) $L_s$ vs.  rotation period
$P$ for 1 M$_\odot$ solar analogs; (b) scatter plot $L_s/L_{bol}$ vs.
Rossby number for open clusters and field stars; and (c) boxplot of
$L_s/L_{bol}$ vs. Rossby number for open clusters and field stars.  The
lines here show the X-ray/rotation correlation (right), saturated
(middle) and supersaturated (left) regimes.  See \S \ref{main_seq.sec}
for references and \S  \ref{sample.sec} for a description of the
boxplot.
\label{main_sequence.fig}}
\end{figure}

For later comparison with pre-main sequence Orion stars, Figure
\ref{main_sequence.fig} shows two results from these studies.  Panel
(a) shows a sample of nearby $\simeq$1 M$_\odot$ field solar analogs,
most with ages between 0.3 and several Gyr.  The soft X-ray emission
closely follows the relation $\log L_s = 31.1 - 2.64 \log P$ erg
s$^{-1}$ where P is the period in days \citep{Gudel97, Gaidos98}.
Figure \ref{main_sequence.fig}b shows the relation between X-ray
emissivity and Rossby number from many $ROSAT$ studies of cluster and
field stars \citep[][kindly updated by S. Randich]{Randich00}.  The
lines indicate three regimes \citep{Randich96}:  \begin{enumerate}

\item For slowly rotating stars, X-ray emission is approximately
linearly dependent on Rossby number as $\log L_s/L_{bol} = -5.0  - 2.1
\log Ro$.

\item Below $\log Ro \simeq -0.8$, main sequence stars exhibit a
`saturated' X-ray level of $\log L_s/L_{bol} = -3.0$.  Saturation is
well-established for several tracers of magnetic activity in several
classes of magnetically active stars \citep{Vilhu87, Fleming95,
Krishnamurthi98}.  Considered together, all manifestations of surface
magnetic fields should not exceed $\sim 1$\% $L_{bol}$, a general limit
on the mechanical power in convection \citep{Mullan84}.  But other
saturation processes may also be involved such as: a limit of field
generation capacity of the underlying dynamo, complete coverage of the
surface by strong fields (unity filling factor of photometric
starspots), or centrifugal forces on large magnetic loops in rapidly
rotating coronae \citep{Randich98, Jardine99}.

\item The most rapidly rotating stars with $P < 0.5$ d lie in a
`supersaturated' regime where X-ray emission drops several-fold below
the saturation limit.  Cluster `ultrafast rotators' with $v~sini \simeq
100-200$ km s$^{-1}$, rotationally coupled W UMa binary stars, and some
dM stars exhibit supersaturation.  Again the cause of the diminution of
activity is uncertain:  perhaps magnetic flux is concentrated toward
the poles, centrifugal forces limit the coronal extent, or coronal
temperatures lie out of the narrow $ROSAT$ passband in these rapidly
rotating stars \citep{Randich98, James00, Stepien01, Mullan01}.

\end{enumerate} 

Despite these interpretational difficulties and some discrepancies
between different samples, the overall agreement over 3.5 orders of
magnitude of X-ray luminosity seen in Figure \ref{main_sequence.fig}b
is probably the clearest empirical indicator of the underlying
relationship between magnetic activity and stellar angular momentum
\citep{Krishnamurthi98}.  In particular, the dependence of
$L_s/L_{bol}$ on mass appears to be relatively weak in main sequence
stars in contrast to the findings we report here for PMS stars (\S
\ref{mass.sec}).

\subsubsection{dM and giant stars \label{dM_giant.sec}}

The $\alpha-\Omega$ dynamo model is less convincing for stars with very
deep convective zones such as M-type dwarfs and post-main sequence
giants;  for these stars, the activity-rotation relation is confusing
and poorly understood. This departure from solar-type main sequence
stars is particularly relevant to PMS stars which are fully convective
at the birthline and (except for very low mass stars) develop radiative
cores as they descend the Hayashi tracks.

Standard interiors models indicate that the convective zone thickens as
mass decreases on the main sequence and the stars become fully
convective below mass $0.3-0.4$ M$_\odot$ (M3$-$M4). Yet, no change
either in the distribution of rotational velocities or the
activity/rotation relation is seen around this spectral type
\citep{Delfosse98}.  This may be explained by deficiencies in standard
interiors models that neglect to consider how magnetic fields can
suppress the onset of complete convection down to $\simeq 0.1$
M$_\odot$ \citep{Mullan01}. There may be a subset of M dwarfs where the
surface activity does not depend on rotation; these may be cases where
the fields are generated throughout the convection zone.  The
rotational evolution of dM stars may be simpler than for higher mass
stars as there is less opportunity for internal redistribution of
angular momentum \citep{Sills00}.

Considerable study has been made of magnetic activity of giants with
masses $1 < M < 3$ M$_\odot$ and bolometric luminosities $3 < L_{bol} <
100$ L$_\odot$ lying at the base of the red giant branch after crossing
the Hertzsprung gap, occupying the same region of the HR diagram as
$<1$ Myr T Tauri stars.  Their interiors range from nearly fully
radiative G giants to K giants with an outer convective zone occupying
$90$\% of the stellar radius.  The strongest effect among these stars
is the `coronal dividing line':  giants with spectral types hotter than
about K1 typically exhibit $\log L_x \sim 28$ to 30 erg s$^{-1}$ ($\log
L_x/L_{bol} \sim -7$ to -5) while cooler giants are usually X-ray
inactive, sometimes with $\log L_x/L_{bol} \leq -10$
\cite[e.g.][]{Ayres81, Hunsch96a, Gondoin99}.

While a rough link between X-ray luminosity and rotation is present
because both are low for the cooler giants, the X-ray/rotation diagram
for the hotter giants shows mostly scatter, up to three orders of
magnitude in $L_x$ for a given rotational velocity \citep{Gondoin99,
Pizzolato00}.  Several stars are known with slow rotation ($v~sini
\simeq 1-3$ km s$^{-1}$) and high X-ray luminosities ($\log L_x \sim 29
- 30.5$ erg s$^{-1}$). A weak X-ray/rotation correlation may be present
for the lower mass ($1.0 < M < 1.5$ M$_\odot$) giants, but an {\it
anticorrelation} between $L_x$ and $v~sini$ may be present among higher
mass ($1.5 < M < 3.0$ M$_\odot$) giants.  These authors suggest that
the strength of the dynamo in these more massive giants is regulated
more by internal differential rotation than the rotation itself.
Computations indicate that turbulence-induced differential rotation
arises as the convective envelope thickens \citep{Kitchatinov99}.
However, it is possible that the coronal dividing line arise from
differences in magnetic field configurations at the stellar surface
rather than differences in dynamo processes \citep{Rosner95}.  A
valuable but inconclusive discussion on issues concerning magnetic
activity in red giants appears in \citet{Strassmeier98}.

\subsubsection{Pre-main sequence stars \label{pms.sec}}

High levels of X-ray emission are ubiquitous among PMS stars, with the
X-ray luminosity function extending from $<10^{28}$ to $10^{31}$ erg
s$^{-1}$ \citep[see review by][]{Feigelson99}.  This is far above
typical main sequence levels of $10^{26}-10^{29}$ erg s${-1}$ but,
because their surface areas are greater, their surface fluxes are
typically an order of magnitude below main sequence saturation levels.
The emission is characterized by high temperatures ($kT \simeq 2$ keV
is typical but 5 to $>10$ keV values are not uncommon; F02a), too hot
to be produced by an accretion shock.  The X-ray emission is usually
strongly variable; for example, the $Chandra$ dataset studied here
indicates that solar mass ONC stars exhibit flares with $L_t(peak) \geq
10^{29}$ erg s$^{-1}$ every few days (F02b). The emission is thus
dominated by flares rather than a soft-spectrum, quiescent corona. The
geometry of the reconnecting fields responsible for the flares is quite
uncertain.  Possibilities include field lines rooted in the stellar
surface as in older stars, field lines extending from the star to the
disk, and fields in a disk corona.

The relationship between activity and rotation for PMS stars is not
well-established.  Although elevated X-ray emission is present during
all PMS phases, rotation is more easily measured during the later
phases when the continuum and sometimes broad emission line excesses of the
`classical' T Tauri phase have subsided.  Most of the measured periods
are obtained from photometric time series of rotationally modulated
cool starspots on `weak-lined' T Tauri stars which are no longer
interacting with their circumstellar disks \citep[e.g.][]{Herbst02}.  A
handful of bright T Tauri stars also have surface Doppler images
\cite[e.g.][]{Donati99, Granzer00} and Zeeman magnetic field
measurements \citep{JohnsKrull99}.

X-ray/rotation studies have concentrated on T Tauri stars in the
Taurus-Auriga complex ($d \simeq 140$ pc), which are often well-studied
and not heavily obscured. Promising evidence for a solar-type dynamo
emerged from the $Einstein$ $Observatory$ when \citet{Bouvier90}
reported an anti-correlation between $F_s = L_s/4 \pi R_\star^2$ and
rotation period in a sample of 13 classical and 8 weak-lined T Tauri
stars.  Their X-ray activity is elevated several-fold above active main
sequence stars with similar rotations.  However, the correlation is
weaker and the scatter greater when a larger $Einstein$ sample of 50
Taurus-Auriga stars are considered \citep{Damiani95}.  Studies of the
entire Taurus-Auriga region with the $ROSAT$ All-Sky Survey gave large
samples showing apparent correlations between X-ray luminosities and
rotational periods and surface velocities \citep{Neuhauser95,
Wichmann00, Stelzer01}.  These results will be discussed with respect
to our findings in \S \ref{X_period.sec}.

The X-ray/rotation relation has also been sought in other nearby star
forming regions.  $ROSAT$ studies of the Chamaeleon I cloud and the
ONC, for example, show most stars lying below the saturation level
without an evident X-ray/rotation correlation \citep{Feigelson93,
Gagne95}.  Two $ROSAT$ samples selected for unusually strong X-ray
emission similarly show no X-ray/rotation correlation, with several
stars overluminous in X-rays compared to saturated main sequence stars
\citep{Preibisch97, Alcala00}.

In summary, a broad correlation with rotational speed is present in
some samples, but considerable scatter is present and the relationship
may not be the same as seen in main sequence stars (Figure
\ref{main_sequence.fig}).  Note, however, that previous investigations
generally had samples too small to permit study of the rotational
effects on X-ray activity independent of other properties such as
stellar mass\footnote{ We do not address here the complex and poorly
understood astrophysics of the rotational evolution of PMS stars.
Possible stages include: spinup during the star formation process when
accretion from the cloud envelope dominates; spindown due to magnetic
coupling between the star and disk; spinup due to angular momentum
conservation as the star descends the Hayashi track; and spindown
during passage to the main sequence, due either to braking by a
magnetic stellar wind or redistribution of angular momentum between the
core and envelope \citep[e.g.][]{Bodenheimer95, Bouvier97b, Stassun99,
Barnes01, Tinker02}.}.

\subsection{Theoretical considerations \label{theor_rev.sec}}

The standard dynamo theory developed for the solar interior and applied
to main sequence and giant stars as outlined above cannot be readily
applied to fully convective stars, as it assumes the field is generated
and amplified at the interface, or tachocline, between the convective
and radiative zones. However, models have been developed where dynamos
operate throughout a convection zone \citep{Durney93}.  If sufficiently
efficient, such a distributed dynamo could not only explain surface
magnetic activity, but could have a considerable effect on the bulk
stellar properties.  For example, a field with 3\% of the energy
density of the gas distributed throughout the interior of PMS stars
shifts the Hayashi tracks several hundred degrees towards the red
compared to standard tracks in the HR diagram  \citep{DAntona00}.

\subsubsection{$\alpha-\Omega$ solar-type dynamo
\label{alpha_omega.sec}}

In a modern dynamo theory for Sun-like stars \citep[e.g.][]{Parker93,
Charbonneau97, Markiel99}, a toroidal field is generated by strong
differential rotation that arises in the thin overshoot layer or
tachocline between the radiative and convective zones (the $\Omega$
effect). These fields are then twisted and transported through the
rotating convective zone to the surface (the $\alpha$ effect). With an
appropriate choice of $\alpha$, such models explain many
characteristics of solar activity including the 22-year cycle, the
`butterfly diagram' of active region magnetic orientations, and
differential rotation in the solar interior inferred from inversion of
helioseismological data \citep[e.g.][]{Charbonneau99}.

For dynamo mechanisms that scale with the Rossby number, the deep
convective zones of PMS stars lead to $\tau_c$ values an order of
magnitude longer than in main sequence stars, giving smaller $Ro$
values and more magnetic field generation at a given rotational period
compared to main sequence stars.  However, the relevance of $Ro$ for
PMS magnetic field generation is not clear.  For example,
\citet{Durney82} suggest that for a distributed dynamo, the efficiency
scales with the depth of the convective region as well as the inverse
of the Rossby number. 

Two detailed calculations of the convective turnover time $\tau_c$, and
hence Rossby numbers, for PMS stars have been reported.  First,
\citet{Gilliland86} considered nonrotating PMS interiord and finds
$\tau_c$ is $\sim 200$ days for fully convective PMS stars at the top
of the Hayashi track.  In higher mass stars, $\tau_c$ drops sharply by
several orders of magnitudes in $\simeq 1$ (10) Myr for $M = 3$
M$_\odot$ (1 M$_\odot$) stars. In lower mass $0.5-1$ M$_\odot$ stars,
$\tau_c$ falls only gradually over $10^7-10^8$ yr.  Second,
\citet{Kim96} provide calculations of $\tau_c$ using updated OPAL
opacities, realistic surface boundary conditions, improved models of
diffusion and rotational mixing, and angular momentum loss by a
magnetized stellar wind.  They treat fully convective Hayashi track
stars with masses between 0.5 and 1.2 M$_\odot$ undergoing solid body
rotation with equatorial surface velocity of 30 km s$^{-1}$
(corresponding to a period $P \simeq 5$ days if $R_* = 3$ R$_\odot$).
Surface rotation is assumed to decay with age as $t ^{-1/2}$ (which may
often not be correct).  They find that $\tau_c$ rises from around 600
to $\ge 1000$ days over several million years in 0.5-1 M$_\odot$ stars,
whereafter it drops to shorter timescales.  More massive $1.0-1.2$
M$_\odot$ stars start at $\tau_c \simeq 700-400$ days and only show the
decline.  This implies that dynamo efficiency is constant (for
solar-mass) or grows $1-2$ orders of magnitude (for sub-solar mass)
stars during the first $\sim 10$ Myr, whereafter it drops by several
orders of magnitude over gigayear timescales.  

We use $\tau_c$ values from \citet{Kim96} in deriving $Ro$ values for
ONC stars below.  We caution that the calculations of $\tau_c$ by
\citet{Gilliland86} and \citet{Kim96} differ both in qualitative
behavior and quantitatively by factors of $2-5$ over the age range of
interest, and even the relevance of the Rossby number for magnetic
field generation or surface magnetic activity in these stars is
uncertain.

\subsubsection{Distributed dynamos \label{dist_dyn.sec}}

A distributed dynamo due to turbulence in the convection zone was first
discussed in detail by \citet{Durney93}. They emphasize that the
turbulent velocity field in a convection zone will generate small-scale
magnetic fields that can attain energy densities comparable to the
kinetic energy density of convective motions.   Rotation may enhance
the rate of field generation but is not essential to the process.  The
principal result of adding an $\Omega$ effect from the boundary between
a convection zone and a radiative core is to build significant energy
densities in large-scale fields, such as those that dominate the solar
cycle.  They argue that small-scale turbulent fields may coexist with
large-scale $\alpha-\Omega$ fields generated in the tachocline, and
should dominate the large-scale fields in stars with deep convective
zones.

Recent calculations have been made of fully convective T Tauri stars
rotating nearly as a solid body with differential rotation around 1\%,
both radially within the convection zone and latitudinally along the
surface \citep{Kuker97, Kitchatinov99, Kuker01}.  Field amplification
occurs throughout the convection zone, and little dependence on bulk
rotation is expected. In other models of PMS interiors, magnetic
activity is inferred to arise from $\alpha-\alpha$ processes, producing
non-axisymmetric and steady fields, in contrast to $\alpha-\Omega$
fields which are typically axisymmetric and oscillatory \citep{Moss96,
Kuker99, Kitchatinov01b}.

\citet[][p. 183f]{Schrijver00} outline a related dynamo concept for stars
with deep convective envelopes. At the base of the convective zones
where the Alfv\'en velocity is low, magnetic fields are subject to
little buoyancy and reside in the same region for a long time.  They
are then wound up and greatly strengthened by differential rotation,
giving a strong field layer analogous to the tachocline in solar-type
stars from which an $\alpha-\Omega$ dynamo can be sustained.

\citet{Mullan01} give a valuable discussion concerning whether a sharp
change in X-ray emission is expected in a star (or ensemble of stars)
that passes from a core-convection zone structure to a completely
convective structure.  No clear prediction can be made:  turning off an
efficient $\alpha-\Omega$ dynamo should reduce the X-ray emission, but
the less efficient $\alpha-\alpha$ dynamo may compensate by operating
over a larger volume.  

Finally, we note that distributed dynamo theories refer to field
generation in the stellar interior and do not specify how these field
emerge onto the surface to produce the extremely large starspots and
violent X-ray flares observed in PMS stars.  A critical issue is
whether the surface magnetic saturation level, as measured by
$L_x/L_{bol}$, could be substantially lower for a distributed dynamo
than a main sequence $\alpha-\Omega$ dynamo.

\subsubsection{Relic and core magnetic fields \label{relic.sec}}

It is possible that the dominant source of magnetic flux in T Tauri
stars are `fossil fields' inherited from the star formation process
rather than generated by a dynamo \citep{Mestel99}.  Poloidal magnetic
fields of order $10^4$ G are roughly expected from compression of
interstellar cloud fields \citep{Dudorov89, Levy91}.  In a fully 
convective PMS star, this fossil interstellar field should
quickly decay due to turbulent magnetic diffusivity.  However, it is
possible that the field may collect into flux ropes which would
resist turbulent diffusion until a radiative core develops \citep{Moss02}. 

PMS magnetic fields might also arise in the radiative core (which forms
at $t \simeq 2$ Myr for a 1 M$_\odot$ star) by capturing flux from the
convective zone.  Such core fields could persist unchanged for billions
of years and could coexist with convective zone dynamo-generated fields
\citep{Tayler87, Moss96, Kitchatinov01a}.  Relic fields trapped in the
larger radiative cores of intermediate mass stars may account for the
high surface fields in Am/Ap stars \citep{Mullan73, Stepien00}.  Unlike
dynamo generated fields, relic fields are likely to have a global
dipole component and may be non-axisymmetric \citep{Kitchatinov01b}.  A
global dipole is needed to produce the large-scale field lines thought
to link the T Tauri star to the circumstellar disk at the corotation
radius \citep[e.g.][]{Hartmann98}.

\subsubsection{Disk-related fields \label{disk_fields.sec}}

T Tauri stars differ from older late-type stars in that they often have
a circumstellar disk.  While the disk is thermodynamically cold and
neutral, sufficient X-rays and cosmic rays likely penetrate and ionize
the disk to freeze in magnetic fields and initiate MHD instabilities
and dynamo processes \citep{Glassgold00}.  Some forms of magnetic
activity, such as the reconnection flares that dominate the X-ray
emission, may thus arise in three locations: at the stellar surface as
in other late-type stars; at the corotation interface between
large-scale dipolar stellar fields and the inner disk \citep{Shu97,
Montmerle00, Birk00}; or above the disk in a magnetically active corona
\citep[e.g.][]{Levy89, Romanova98, Merloni01}.  There is a wealth of
evidence for strong activity at the stellar surface, but the strong
fluorescent 6.4 keV iron line seen in two protostars \citep{Koyama96,
Imanishi01} may be evidence that X-ray flares occur in close proximity
to the disk. This issue of the geometry of reconnecting magnetic field
lines in T Tauri systems is discussed in detail by F02b.

\section{The X-ray data \label{data.sec}}

\subsection{Observations}

The Orion Nebula Cluster (ONC) is the richest young star cluster within
500 pc with $\simeq 2000$ members concentrated in a 1 pc ($8^\prime$)
radius sphere \citep{Odell01}.  The full initial mass function from a
45 M$_\odot$ O star to dozens of substellar brown dwarfs is present.
Over 1500 stars are not deeply embedded and have $V<20$ magnitudes,
$\sim 1000$ of which have high-quality photometry and spectroscopy
\citep[][and subsequent updates to the database]{Hillenbrand97}.  This
gives locations on the HR diagram from which stellar ages and masses
are inferred from theoretical stellar interior models
\citep{DAntona97}.  We ignore here the X-ray population of deeply
embedded stars which lies behind the ONC around the OMC 1 cloud cores.

The ONC was observed with the ACIS-I imaging array on board {\it
Chandra} twice during the inaugural year of the satellite, on 12 Oct
1999 and 1 Apr 2000, for $\simeq 12$ hours on each occasion.  The
satellite and instrument are described by \citet{Weisskopf02}.  The
reader should consult F02a for an atlas of the field, full description
of the data reduction procedures, and properties of the 1075 X-ray
sources found in the field.

\subsection{Sample and database \label{sample.sec}}

Of the 1075 ACIS ONC sources, we consider stars with estimated ages and
masses \citep{Hillenbrand97} and further eliminate stars with $M > 3$
M$_\odot$\footnote{For intermediate- and high-mass ONC stars with $M >
3$ M$_\odot$, it is not clear that the X-rays arise from the optically
characterized star rather than from unseen companions (F02a, \S
5.1-5.2).  Only one of these omitted stars has a measured rotation
period: the B8 star JW 660 with period of 6.15 days and a high X-ray
luminosity of $\log L_t = 31.1$ erg s$^{-1}$ ($0.5-8$ keV band).}  The
resulting sample of 525 stars is listed in Table \ref{srcs.tab}.
Absorption is not large for most of these stars: 47\% have $A_V \leq
1$, 95\% have $A_V < 5$, and for 77\% the difference between the
observed total band ($\log L_t$) and absorption-corrected ($\log L_c$)
X-ray luminosities does not exceed 0.3.  The $\log L_t$ values in the
$0.5-8$ keV band thus reflect the true emission with reasonable
accuracy.  The $\log L_s$ luminosities in the soft $0.5-2$ keV band
will be more seriously affected by absorption, and are provided only to
permit comparison with earlier $ROSAT$ soft band results.  Note that
the main source of scatter in the X-ray luminosities is the intrinsic
variability of the sources during the two observations.

Table \ref{srcs.tab} gives: the ACIS-I CXOONC source name (column 1);
associated optical star \citep[column 2, most are designated JW
from][]{Jones88}; stellar bolometric luminosity, mass and age (columns
$3-5$);  a circumstellar disk indicator (column 6); rotational period
with reference (columns $7-8$); estimated Rossby number (column 9);
soft and total band X-ray luminosities (columns $10-11$), and the ratio
of total band X-ray to bolometric luminosity (column 12). Columns $1-5$
and $10-11$ are extracted from Tables 2 and 3 of F02a.  As in F02b, we
considered stellar ages below $\log t = 5.5$ yr to be upper limits
because of difficulties in establishing the zero-age point in
evolutionary calculations \citep[e.g.][]{Wuchterl01}.  The disk
indicator is based on the criteria given by F02b with data from F02a.
A {\bf +} symbol indicates a near-infrared photometric excess
$\Delta(I-K) > 0.3$ and/or association with a Herbig-Haro outflow,
far-infrared source or imaged proplyd; a {\bf $-$} symbol indicates
$\Delta(I-K) < 0.3$ and no association of these types; and $\ldots$
indicates insufficient information for classification.  The $\Delta(I-K)$
measurements are from \citet{Hillenbrand98}. 

The photometric rotational periods are extracted from Table 2 of F02a.
The code for rotation period references is: C = \citet{Carpenter01}; H
= \citet{Herbst00} and \citet{Herbst02}; and S = \citet{Stassun99}.  A
few rotation periods have been updated from those given in F02a based
on the final results of \citet{Herbst02}, and stars with discrepant
reported photometric periods are listed in the Notes to Table 2 of
F02a.  We do not supplement these with 43 new periods estimated from
the projected Doppler surface velocity measured spectroscopically by
\citet{Rhode01}.  Periods derived from spectroscopy are inaccurate due
to the unknown inclinations of individual stars, and a systematic
overestimation compared to photometric periods is present.

Column 9 of Table \ref{srcs.tab} lists Rossby numbers $Ro$ derived from
the observed rotation periods and $\tau_c$ estimated from Figure 3 of
\citet{Kim96} in the $0.5-1.2$ M$_\odot$ range (\S
\ref{alpha_omega.sec}).  Due to these restrictions, only 36 values are
given.  

Columns (10-12) give the X-ray luminosities $\log L_s$ (erg s$^{-1}$)
in the soft $0.5-2$ keV band, $\log L_t$ in the total $0.5-8$ keV band,
and the ratio $\log L_t/L_{bol}$ where $L_{bol}$ is obtained from
\citet{Hillenbrand97}.  The $\log L_s$ and $\log L_t$ values are
obtained from Table 3 of F02a; see their \S 2.6-2.9 for
details.\footnote{We provide $\log L_s$ values to facilitate comparison
of the $Chandra$ results to earlier $ROSAT$ results.  When comparing
PMS to main sequence X-ray emissivities, recall that the
$ROSAT$-derived $\log L_s/L_{bol}$ values for main sequence populations
are systematically lower than our $\log L_t/L_{bol}$ value due to our
wider bandwidth ($0.5-2$ keV for $L_s$ $vs.$ $0.5-8$ keV for $L_t$).
For typical PMS spectra, $L_t$ values are typically a factor of 2
higher than $L_s$ values due to this bandwidth effect, and may be
higher yet due to interstellar attenuation of $L_s$.  In particular, we
note that the $ROSAT$-derived main sequence saturation level $\log
L_s/L_{bol} = -3.0$ (Figure \ref{main_sequence.fig}b) is equivalent to
about $\log L_t/L_{bol} \simeq -2.7$ for lightly absorbed stars.}.

In \S \ref{depend.sec}-\ref{rotation.sec}, we visualize the data from
Table \ref{srcs.tab} using boxplots in addition to two-dimensional
scatter plots.  Boxplots are a simple nonparametric graphical tool for
visualizing and comparing univariate distributions widely used in many
fields \citep{Tukey77, McGill78}. The center of the box indicates the
median value and the `hinges' (ends) of the box enclose the 25\% and
75\% quartiles of the data.  `Whiskers' (error bars) extend from the
box to the largest data value less than 1.5 times the quartile range.
Circles show outliers if present;  for a Gaussian distribution, about 1
in 100 points will be an outlier.  If the `notches' (indented regions
around the medians) of two boxes on the same plot do not overlap, then
the two population medians are different with $>95$\% confidence based
on an assumption of asymptotic normality of the standard deviation of
the medians (i.e. large-N samples).  The width of the boxes is scaled
to the square root of the number of points included in each box so that
the wider boxes have greater statistical reliability than narrower
boxes.  The range of each box along the abscissa was chosen by us in an
arbitrary manner. The graphics were produced with {\bf R} \citep{Ihaka96},
a public-domain statistical software package closely related to the
commercial {\bf S-Plus} package.  {\bf R} software and documentation
can be obtained at \url{http://www.r-project.org}.

\subsection{Sample completeness \label{complete.sec}}

Although Table \ref{srcs.tab} is by far the largest dataset of magnetic
activity measurements for PMS stars with measured stellar properties,
we must consider systematic biases present in the sample:
\begin{enumerate}

\item Our sample is first restricted to 979 ONC stars placed on the HR
diagram lying within the ACIS field.  This sample is estimated to be
100\% complete for all ONC stars with $M \geq 0.5$ M$_\odot$ with $A_V
\leq 0$, and for $A_V < 2.5$ 100\% complete for $M \geq 1$ M$_\odot$
and $50-70$\% complete above the substellar limit \citep[][, \S
4.3]{Hillenbrand97}.  The main omission are very-low-mass M stars and
brown dwarfs which show up in deep K-band studies \citep{Hillenbrand00}.

\item Of these 979 stars, our sample is restricted to 525 stars
detected with $Chandra$ having ACIS count rates above $0.1-0.4$ cts
ks$^{-1}$ in the $0.5-8$ keV band, where the higher values are due to
reduced sensitivities from the poor point spread function towards the
outer portions of the cluster (F02a, \S 2.12). For most cluster members
with typical intrinsic PMS X-ray spectra and low absorptions, this
limit corresponds to $\log L_t = 28.0-28.5$ erg s$^{-1}$ although some
limits reach $\log L_t = 29.0$ erg s$^{-1}$.  Here also a strong bias
in mass is present: $\simeq 90$\% (F02a, \S 5.2) of ONC members with
$M>1.5$ M$_\odot$ are present compared to roughly 25\% of PMS brown
dwarfs (F02a, \S 5.6).

\item Of these 525 stars, 232 have measured photometric periodicities
interpreted as rotationally modulated starspots.  By comparing
spectroscopically measured $v~sin~i$ rotational velocities for ONC
stars with and without detected photometric starspots, \citet{Rhode01}
have found that the stars with modulated starspots have the same
rotational distribution as the underlying ONC population.  Also, the
latest study of \citet{Herbst02}, which provides most of the
photometric rotation periods used here, extends period measurements
down to $M \simeq 0.1$ M$_\odot$.  The rotation measurements should
thus not contribute any further bias to our sample except below $M
\simeq 0.1$ M$_\odot$.

\item Both the optical and X-ray data have arcsecond ($\sim 500$ AU)
resolution and thus see the majority of binary and multiple systems as
a single star \citep{Mathieu94}.  We assume that both the optical and
and X-ray light is dominated by a single primary component.  This
assumption also tends to deemphasize the presence of lower mass stars
from our sample.  

\end{enumerate}

We conclude that the principal bias in our sample of 525 stars involves
stellar mass and associated variables such as bolometric luminosity.  A
double bias is present: the underlying optical sample is deficient in
low mass stars compared to the underlying cluster, and the X-ray
observation is deficient in detecting these stars.  A more complete
sample would thus have many more objects at low masses with
characteristically lower X-ray luminosities.  The bias is nearly absent
for masses $0.7 < M < 3$ M$_\odot$.  From Table 5 in F02a and Table 1
in F02b, we find there are only four\footnote{We omit JW~991 because of
its low probability of cluster membership \citep{Jones88}.} undetected
stars in this mass range: P 1892 with $M = 2.6$ M$_\odot$, JW~531 with
$M = 2.5$ M$_\odot$, JW~608 with $M = 1.8$ M$_\odot$ and JW~62 with $M
= 1.4$ M$_\odot$.  These stars are shown as arrows in the scatter plots
below.

\subsection{Sources of uncertainty \label{uncertainty}}

As considerable scatter appears in the correlation plots presented
below, it is important to discriminate the degree to which these arise
from measurement errors or from true astrophysical variance.  The broad
band $0.5-8$ keV X-ray luminosities $\log L_t$ in most cases have
rather small ($\Delta \log L_t = \pm 0.1$) statistical uncertainties,
but the intrinsic variability due to X-ray flaring is frequently
$\Delta L_t = \pm 0.3$ during the two 12-hour $Chandra$ observations
(F02a, \S 2.9) and sometimes exceeds 1.0 (F02b). The long-term
variability of a star will obviously exceed the variability found
during the limited observations available here.  We thus expect all
samples of PMS stars to exhibit significant scatter in X-ray
luminosity, roughly $\Delta \log L_t = \pm 0.5$ for the majority of
stars, due to statistics and variability.

Uncertainty or systematic errors may also be present in other stellar
parameters.  $\log L_{bol}$ is relatively well-established with errors
about $\pm 0.15$ by the photometry and spectrometry of
\citet{Hillenbrand97}.  Stellar masses and ages depend on the model
assumptions of the evolutionary tracks adopted in our study
\citep{DAntona97}.  These quantities will systematically change with
differing assumptions regarding the equation of state, mixing length
theory, accretion, rotation, and the internal magnetic field
\citep{DAntona00, Palla01}.  The effects of even modest observational
error on parameters derived from evolutionary tracks, especially
stellar age, may be significant: an uncertainty $\Delta T_{eff} = \pm
100$ K and $\Delta L_{bol} = \pm 0.1$ produces fractional errors around
$\Delta \log M = \pm 0.1$\% and $\Delta \log t = \pm 0.5$
\citep{Siess01}.  Rotational periods generally have almost no
statistical uncertainty but sometimes suffer large errors if the wrong
peak in a periodogram is chosen.  A few stars in our sample with
discrepant reported periods of this type are listed in the notes to
Table 1 of F02a.

We thus expect scatter in various stellar properties, particularly age,
due to observational error, plus possible systematic errors in
properties due to model assumptions.  In most cases, the latter may
produce offsets or stretching of the plotted axes, but will not affect
overall strength of a correlation.  The greatest danger would arise if
both the X-ray luminosity and another property of interest were
mutually dependent on magnetic field generation, producing spurious
correlations.  However, this problem does not appear to be present: PMS
model interiors with magnetic fields tends to have cooler surfaces
which would yield lower inferred masses \citep{DAntona00}.  In
contrast, we find below (\S 4.3) that Orion stars with stronger
magnetic activity have higher rather than lower masses than those with
weak activity.

\section{X-ray dependencies on stellar properties \label{depend.sec}}

We present here empirical results relating the X-ray emission, viewed
as an indicator of magnetic activity, to the bulk properties of the ONC
PMS stars: bolometric luminosity, mass, age, presence of disk, and
surface rotation.  In some cases we elucidate longstanding
relationships found from past studies (\S \ref{pms.sec}), while in
other cases we reveal new phenomenology.  The findings are summarized
in \S \ref{summary.sec}.

\subsection{X-ray and bolometric luminosities \label{lbol.sec}}

With a sample population far larger than previously available, we can
now see why a correlation between $L_s$ and $L_{bol}$ has been seen in
past studies of PMS stellar populations but with inconsistent
quantitative results \citep[e.g.][]{Walter81, Feigelson93, Casanova95,
Gagne95, Stelzer01, Preibisch02, Getman02}.  Figure \ref{Lx_Lbol.fig}a
shows a broad correlation over three orders of magnitude, roughly
consistent with the linear relationship $\log L_t \simeq 29.8 + \log
L_{bol}$ erg s$^{-1}$ or, as seen in Figure \ref{Lx_M.fig}c,  $\log
L_t/L_{bol} \simeq -3.8$.  But, due to the selection bias against
X-ray-faint low-mass stars (\S \ref{complete.sec}), it is likely that
the median X-ray luminosity at low $L_{bol}$ values is overestimated
here, leading to a steeper true relation.  For example, the data could
be modelled as $L_t \propto L_{bol}^2$ with a saturation limit at high
luminosities.  Although difficult to quantify due to the scatter and
bias, examination of the notches in the boxplot (Figure
\ref{Lx_Lbol.fig}b) shows that the overall correlation has very high
statistical significance.

Whatever the underlying relationship between X-ray luminosity and
$L_{bol}$, a great deal of scatter is present\footnote{The outliers
with high $L_{bol}$ and very low $L_t$ are discussed in \S
\ref{mass.sec}; they are most vividly seen in Figure \ref{Lx_M.fig}c.
Similar outliers are found by \citet{Preibisch02} from $Chandra$
observations of the IC 348 young stellar cluster.}. At any given
$L_{bol}$ value, the dispersion in $L_t$ or $L_t/L_{bol}$ is such that
half of the stars lie greater than a factor of 3 away from the
predicted value, and some are discrepant by more than an order of
magnitude.  This dispersion must be astrophysical in origin as it is
too large to arise from observational or absorption effects. It is
likely that X-ray flaring is the main contributor to this dispersion,
but other dependencies on other variables may also be important.

\subsection{X-ray emission and stellar size \label{size.sec}}

As most T Tauri stars have similar surface temperatures $T_{eff}$,
bolometric luminosity is closely related to stellar surface area via
$L_{bol} = 4 \pi R^2 \sigma T^4_{eff}$ where $\sigma$ is the
Stefan-Boltzman constant.  X-ray luminosities which scale with
$L_{bol}$ will therefore also scale with stellar surface area, radius
and volume.  Figure \ref{Lx_V.fig} shows one of these relationships:
X-ray emission compared with stellar volume in units of $V_\odot = 4 \pi
R^3_\odot/3$.  Recognizing that the median levels for the smallest
stars is probably overestimated due to undetections (\S
\ref{complete.sec}), we find that X-ray luminosity scales roughly as
$L_t \propto V^{2/3} \propto R^2$.

A similar, but considerably steeper, activity-radius effect has been
found in a sample dM 1V main sequence stars by \citet{Houdebine97}.
They find that H$\alpha$, Ca II H\&K and soft X-ray activity indicators
scale with absolute magnitude which, for constant surface temperature,
itself scales with radius, surface area and volume.  Stated another
way, early dMe stars are more luminous, and hence larger, than less
active dM stars.  The dM X-ray sample is small and suggested a
relationship around $L_x \propto R^7$.

\subsection{X-ray emission and stellar mass \label{mass.sec}}

Figure \ref{Lx_M.fig} shows scatter diagrams and boxplots of X-ray
emission as a function of stellar mass.  A comparison of panels (a) and
(b) to those in Figure \ref{Lx_Lbol.fig} shows that mass is as strong a
predictor for X-ray emission as bolometric luminosity, although the
X-ray/mass relation has only occasionally been noticed in past studies
with discrepant quantitative results \citep[e.g.][]{Feigelson93,
Neuhauser95, Preibisch02}.  Given a strong $L_t-L_{bol}$ correlation, a
similar $L_t-M$ relation is expected from a coeval PMS population given
the tilt of the isochrones with respect to the isomass lines in the HR
diagram.  The relationship appears steeper than linear, roughly
consistent with $\log L_t \simeq 30.2 + 1.5 \log M$ erg s$^{-1}$, but
again we recall the selection bias (\S \ref{complete.sec}) that should
increase the slope of this relation at low masses. This is consistent
with the recent $Chandra$-based result $\log L_t = 30.10 + 1.97(\pm
0.24) \log M$ erg s$^{-1}$ derived by \citet{Preibisch02} in the IC 348
young stellar cluster over a mass range similar to that considered
here\footnote{The observed $L_t-$mass correlation may be affected by
unresolved binarity, which is likely to be present in over half of the
ONC `stars' under study \citep{Mathieu94}.  However, it seems unlikely
that the effect is very significant.  If fainter secondary components
were responsible for the X-ray emission, then the low-mass systems
should show as wide a spread in $L_t$ as high-mass systems and the
$L_t$-mass correlation would be weak.  A $ROSAT$ study of nearby T
Tauri stars confirms that the X-ray emission of primaries dominates
over the secondaries in resolved wide binaries \citep{Konig01}.  Note
however that we do believe binarity is important for the interpretation
of X-ray emission from higher mass ($M > 2$ M$_\odot$) stars.}.

The $\log L_t/L_{bol}-M$ diagram (Figure \ref{Lx_M.fig}c) dramatically
reveals an effect distinct from the general $L_t-M$ relationship:
X-ray emission from the higher mass stars in the sample with $2.0 < M <
3.0$ M$_\odot$ has an enormous dispersion.  It is possible that, for $M >
2$ M$_\odot$, the ONC population can be divided into two classes.  The
majority of these $2-3$ M$_\odot$ stars have $-5 < \log L_t/L_{bol} <
-3$ like virtually all lower mass stars, while a
minority\footnote{There is no indication these X-ray-weak stars are
foreground interlopers, as their proper motions have $98-99$\%
probabilities of cluster membership \citep{Hillenbrand97}.  These
stars, however, are older than most ONC stars; it possible that both
mass and age are involved in their unusually low magnetic activity.
Note that weak evidence for a decay in X-ray emission as PMS stars age
was reported for $0.7 < M < 1.4$ M$_\odot$ stars by F02b, and is
discussed again in \S \ref{age.sec} below. \label{evol.foot}} show $-7
< \log L_t/L_{bol} -5$.  The latter low X-ray emissivities are
ubiquitous for the intermediate-mass $3 < M < 30$ M$_\odot$ ONC stars
(see Figure 12a in F02a).  Two interpretations of this difference in
X-ray behavior of intermediate- and low-mass PMS stars are outlined
in \S 6.2.  

\subsection{X-ray emission and circumstellar disks \label{disk.sec}}

From the very beginning of X-ray studies of PMS populations, most
studies found that accretion and outflows associated with `classical' T
Tauri star-disk interactions were not essential ingredients for
elevated X-ray levels.  This is often shown as an absence of
correlation between X-ray and H$\alpha$ emission when a full PMS
population of weak-lined and classical T Tauri stars is treated,
although an X-ray/H$\alpha$ correlation may be present within the
weak-lined T Tauri stars alone where both arise from magnetic activity
\citep[e.g.][]{Montmerle83, Feigelson93, Damiani95, Casanova95,
Gagne95}.  In contrast, some studies find that weak-lined T Tauri stars
(defined by weak H$\alpha$ emission) are an order of magnitude more
X-ray luminous than classical T Tauri stars \citep{Neuhauser95,
Stelzer01}.  But this is likely due to misclassifications and
incompleteness in the sampling of X-ray-faint weak-lined T Tauri stars
in contrast to the good optical sampling of X-ray-faint classical T
Tauri stars \citep{Preibisch02}.

We consider here the photometric near-infrared excess measure
$\Delta(I-K) > 0.3$ as a discriminant of the presence of a disk, which
is not necessarily the same as strong optical emission lines which
indicate the presence of an {\it accreting} disk.  Figure
\ref{Lx_disk.fig} shows no important relationship between X-ray
emission and the presence of a disk. (Another view of this result
appears in the middle panel of Figure 10 in F02a.) Figure
\ref{Lx_disk.fig}c shows that mass, which is a strong correlate of
$L_t$, is not an important confounding variable in this result.

\subsection{X-ray emission and stellar age \label{age.sec}}

Low mass stars evolve in many respects during their descent along the
Hayashi tracks:  the star contracts; brief periods of deuterium and
lithium burning occur; a radiative core forms and grows although most
of the star is convective; and star-disk interaction declines or
terminates, perhaps releasing the star from rotational coupling with
the disk.  While most ONC stars appear to have formed within the past 2
Myr, a tail of stellar ages appears to extend beyond 10 Myr, although
it is not clear these ages are accurate.  Alternatively, the older Myr
stars in the field may be interlopers from the older Orion Ia-c OB
associations \citep[see discussion in][]{Hillenbrand97, Hartmann01}. 

Past study of the evolution of X-ray emission along the Hayashi tracks
has been been limited and somewhat confusing. In $ROSAT$ studies of
individual PMS clusters, \citet{Feigelson93} report a tentative drop of
soft band $L_s$ from $<1$ to 10 Myr while \citet{Neuhauser95} report a
rise with age.  \citet{Kastner97} collect average soft X-ray levels for
stars from several clusters of different ages and find that $<\log
L_s/L_{bol}>$ rises an order of magnitude over tens of Myr.   We
caution that comparisons of mean X-ray luminosities of different
clusters is subject to systematic error due to different X-ray
sensitivities and different levels of prior knowledge of the cluster
memberships.  A rise in X-ray emissivity with PMS age is consistent
with a model of stellar angular momentum evolution where surface
rotation (and presumably the internal magnetic dynamo efficiency) rises
as star-disk rotational coupling ends and the star contracts
\citep{Bouvier97b, Barnes01}.  This model is supported by study of the
$\eta$ Cha cluster, a recently identified older PMS cluster with $t =
9$ Myr stars, where nearly all have unusually short rotational periods
and high X-ray luminosities around $\log L_s/L_{bol} \simeq -3$
\citep{Mamajek00, Lawson01}.

Figure \ref{Lx_age.fig} shows the X-ray/age relationship found for the
ONC sample discussed here.  Recall that ages were estimated from the
evolutionary tracks of \citet{DAntona97} based on the photometry and
spectroscopy of \citet{Hillenbrand97}, and that we truncate all
extremely young inferred ages at 0.3 Myr.  Panels (a) and (b) reveal a
small but statistically significant decline in X-ray luminosity from a
median level of $\log L_t \simeq 29.6$ erg s$^{-1}$ for ages $<1$ Myr
to $\log L_t \simeq 29.2$ erg s$^{-1}$ for ages $> 10$ Myr.  A similar
but steeper drop in $L_t$ is found when the $0.7-1.4$ M$_\odot$ solar
analogs are considered alone (F02b).  We also note that the dispersion
in X-ray luminosities decreases monotonically with age from $>3$ to 2
orders of magnitude.

Panels (c) and (d) show that this fall in X-ray luminosity is roughly
equal to the decrease in bolometric luminosity from 0.3 to 10 Myr, so
that the X-ray emissivity $\log L_t/L_{bol}$ is roughly constant at
-3.8.  But a distinctive change is seen among the oldest ONC stars:
with the exception of a single intermediate mass outlier (see \S
\ref{mass.sec}), all of the 13 ONC stars with apparent ages between 10
and 30 Myr have unusually high X-ray emissivities with $\log
L_t/L_{bol} \simeq -3$ at the main sequence saturation level, similar
to the $\eta$ Cha finding.  There are several possible interpretations
for these stars.  If they are indeed cluster members and are correctly
placed in the HR diagram, they suggest an increase of $L_t/L_{bol}$
with age.  However, if they have been erroneously placed on the
diagram, due perhaps to underestimation of their extinction, then
$L_{bol}$ would be higher and the $L_t/L_{bol}$ ratio consistent with
the bulk of the ONC PMS stars.

\section{X-ray emission and surface rotation \label{rotation.sec}}

The relationships between X-rays and rotation in ONC PMS stars are
shown in Figures \ref{Lx_rot.fig} and \ref{Lx_Rossby.fig}.  They should
be compared to analogous graphs of main sequence stars shown in Figure
\ref{main_sequence.fig} which are discussed in \S \ref{main_seq.sec}.

\subsection{X-rays and rotational period \label{X_period.sec}}

Figures \ref{Lx_rot.fig} and \ref{main_sequence.fig} immediately show
two differences between PMS and main sequence magnetic activity: a
large fraction of ONC stars have considerably stronger X-ray emission
than main sequence with similar rotation periods; and the strong main
sequence anticorrelation between X-rays and period is dramatically
absent in the ONC population\footnote{The locus of ONC stars in Figure
\ref{Lx_rot.fig}c also does not follow the roughly parabolic locus,
peaking around 1 day, seen in dM stars \citep{James00, Mullan01}.}.
Instead, a {\it correlation} in average luminosities with period is
marginally  present (compare the boxplot notches in Figure
\ref{Lx_rot.fig}b) such that stars with periods $P > 10$ days are about
4 times more X-ray luminous on average than stars with $P < 2$ days.
This trend is in the opposite direction of the strong {\it
anticorrelation} seen in main sequence stars, for stars with similar
periods; for example, for solar-mass stars shown in Figure
\ref{main_sequence.fig}a, the X-ray luminosity of stars with $P > 10$
days is $\sim 100$ times smaller than those with $P \simeq 2$ days.
The $\log L_t/L_{bol}$ $vs.$\ $P$ diagram similarly does not show any
sign of the steep decline in X-ray luminosity with period seen in main
sequence stars over a similar period range (compare Figure
\ref{Lx_rot.fig}c with Figure \ref{main_sequence.fig}a).

Perhaps the most challenging characteristic of this finding to explain
are the high X-ray luminosities of very slowly rotating PMS stars.
Such stars had been occasionally found in the past; for example,
\citet{Preibisch97} noted that the ONC star JW 157 (= P 1659) has a
surprisingly high X-ray emissivity $\log L_s \simeq 31.5$ erg s$^{-1}$
for its 17.4 day period, and \citet{Lawson01} find RECX 10 in $\eta$
Chamaeleon has $\log L_s/L_{bol} = -2.9$ erg s$^{-1}$ with $P = 20.0$
days. Both of these are slowly-rotating weak-lined T Tauri stars,
although JW 157 appears to be very young ($\log t < 5.5$ yr) while RECX
10 is old ($\log t = 7.0$ yr).  The ONC provides a sample of $\simeq
30$ such stars with $P > 10$ days and $\log L_t/L_{bol} = -4 \pm 1$
with a wide range of masses.

We recall that some $Einstein$ and $ROSAT$ studies report
X-ray/rotation correlations while others do not (\S \ref{pms.sec}).
Perhaps the clearest case that is discrepant from our result is the
$ROSAT$ study of Taurus-Auriga PMS stars by \citet{Stelzer01}.  They
find that, for 39 stars in the soft X-ray band, X-ray emission
systematically decreases from $\log L_s \simeq 30.6$ to 29.1 erg
s$^{-1}$ and $\log L_s/L_{bol} \simeq -3.0$ to $-4.5$ as rotational
period increases from $\simeq 1$ to 10 days.  We suspect that this
discrepancy arises from incompleteness in the Taurus-Auriga sample; it
is difficult to define and study the population of this large cloud
complex where star formation has occurred in cores dispersed over 500
square degrees.  First, arguments have been put forward that
Taurus-Auriga PMS stellar samples are deficient both in high mass stars
\citep{Walter91} and faint low mass weak-lined T Tauri stars
\citep{Luhman00, Preibisch02}.  The effects of such missing stars on an
X-ray/rotation diagram is unknown.  Second, rotational periods of
Taurus-Auriga stars were typically obtained from photometric
observations of specific PMS stars with observing sessions spanning
$\simeq 10 - 40$ days \citep[e.g.][]{Bouvier86, Bouvier97a} and result
in periods for only 39 of 168 stars detected in the study of
\citet{Stelzer01}.  In contrast, most ONC periods were obtained from
observing runs spanning several months or years \citep{Herbst00,
Herbst02}, and result in periods for 232 of 525 stars in the present
ONC study.  It is thus possible that an improved study of Taurus-Auriga
rotations would show a subpopulation of slow rotators with strong X-ray
emission which would remove the X-ray/rotation correlation found by
\citet{Stelzer01}.

\subsection{X-rays and Rossby number \label{X_Rossby.sec}}

It is well-known that combining stars of different masses can blur
relations between magnetic activity indicators and rotational periods.
We address this in two ways.  First, examination of individual symbols
in the scatter plots in Figure \ref{Lx_rot.fig}, which represent
different mass ranges, shows no evidence of the expected decrease in
X-ray emission with increasing period within individual mass strata.
Second, we consider the X-ray relation to Rossby number, which is very
effective in removing mass-dependent effects in the context of
$\alpha-\Omega$ dynamo models \citep{Noyes84, Montesinos01}.  As
described in \S \ref{sample.sec}, we obtain Rossby numbers from the
convective turnover times for PMS stars calculated by \citet{Kim96},
recognizing that they assume a single rotation rate and are available
only for $0.5-2$ M$_\odot$ stars. The results are shown in Figure
\ref{Lx_Rossby.fig}; panel (c) is most valuable for its comparison with
the main sequence X-ray/Rossby number relation (Figure
\ref{main_sequence.fig}c).

The X-ray/Rossby number plot (Figure \ref{Lx_Rossby.fig}b) gives a
possible explanation for the absence of the expected X-ray/rotation
relation.  Due to the very short calculated convective turnover times
at the base of the deep convection zones of PMS stars, most ONC PMS
stars around $M \sim 1$ M$_\odot$ lie in the supersaturated regime
rather than the linear regime where X-ray emission inversely correlated
with Rossby number.  Extremely long rotation periods around 100 days
would be needed to move the ONC stars into the linear regime.

\section{Discussion \label{disc.sec}}

It is valuable to first recognize why this study may achieve results
not available to previous observations.  For PMS stars, X-rays from
reconnection flares are the most easily observed indicator of surface
magnetic activity.  Optical emission line indicators useful in other
types of stars are often confused by lines due to accreted or ejected
matter, and the ultraviolet is ofen obscured by interstellar matter.
Doppler imaging and Zeeman effect studies are very valuable for mapping
surface fields, but have to date been obtained for only a handful of
the brightest T Tauri stars.  X-ray emission, on the other hand, is
typically elevated $10^{2 \pm 1}$ times above solar levels during all
phases of PMS evolution \citep{Feigelson99}.  PMS spectra show typical
plasma energies around $1-3$ keV and are sometimes dominated by plasmas
as hot as $\sim 10$ keV (F02a), and can therefore been studied even in
the presence of considerable interstellar absorption.  A 2 keV photon
has the same penetrability as a 2 $\mu$m near-infrared photon, and is
comparable to mid-infrared emission above 5 keV \citep{Montmerle02}.
Finally, the ONC provides the largest and best defined PMS sample in
the nearby Galaxy in the sense that virtually all members of the
cluster appear in the optical/infrared sample with very few
contaminants from unrelated objects.  The ONC has the largest sample of
PMS stars with detailed optical photometric, spectroscopic and rotation
measurements.  While nearly all earlier X-ray telescopes studied the
ONC, only {\it Chandra} has the sensitivity and resolution to resolve
the crowded cluster core (except for multiple systems).  Our
observations, for example, achieve more than an order of magnitude
greater sensitivity than $ROSAT$ observations of the ONC.

\subsection{Summary of findings \label{summary.sec}}

In this light, the principal findings from examination of bivariate
relations between X-ray emission and stellar properties for
well-characterized ONC stars are:  \begin{enumerate}

\item X-ray luminosities are strongly correlated with several closely
coupled stellar properties: bolometric luminosities, stellar size
(radius, surface area and volume), and mass (\S
\ref{lbol.sec}-\ref{mass.sec}).  The $\log L_t - \log L_{bol}$
relation, for example, is roughly linear and consistent with an average
$\log L_t/L_{bol} \simeq -3.8$. This is an order of magnitude below the
main sequence saturation level.  The $\log L_t-$size relations are
consistent with X-ray luminosities scaling linearly with stellar
surface area.  The dispersion about the relation is high and can be
largely attributed to X-ray variability and flaring.  The relationship
between X-ray luminosities and mass is steeper than linear, and a sharp
decrease by more than a factor of 10 in X-ray emissivity $\log
L_t/L_{bol}$ is seen in some $2-3$ M$_\odot$ stars.  This drop becomes
ubiquitous for ONC stars with $M > 3$ M$_\odot$.

\item The presence or absence of a circumstellar disk, as measured by
near-infrared photometric excess, appears to have no influence on X-ray
luminosities or emissivities. (\S \ref{disk.sec})

\item  X-ray luminosities shows a mild decline as stars age and descend
the Hayashi track (\S \ref{age.sec}). Because $L_{bol}$ also falls, the
ratio $\log L_t/L_{bol}$ is constant for $t < 10$ Myr and may rise to
the main sequence saturation level during $10 < t < 30$ Myr.

\item Most importantly for our purposes, X-ray luminosities and
emissivities are higher than seen in main sequence stars for any given
rotational period, and show a slight {\it rise} with rotational period
over the range $0.4 \leq P \leq 20$ days in contrast to the strong {\it
decline} seen over the same range in main sequence stars (\S
\ref{X_period.sec}).  However, the result may be consistent with the
main sequence X-ray/Rossby number diagram, as ONC stars appear to lie
in the `supersaturated' regime at low Rossby numbers (\S
\ref{X_Rossby.sec}).

\end{enumerate}

\subsection{Implications for dynamo models}

Clearly PMS stars do not exhibit the standard empirical
activity-rotation relationships seen in main sequence stars attributed
to an $\alpha-\Omega$ dynamo (\S \ref{main_seq.sec}). The X-ray
emission of an ensemble of mass-stratified PMS stars is unaffected by
differences in rotation periods from 0.4 to 20 days, whereas the X-ray
emission of main-sequence stars declines by a factor of $10^3$ over
this same period range\footnote{The comparison between main sequence
and PMS activity may appear somewhat paradoxical at first glance: PMS
X-ray {\it luminosities} ($\log L_t$) are considerably elevated above
main sequence levels, particularly for slow rotators, but PMS X-ray
{\it emissivities} ($\log L_t / \log L_{bol}$) are below the main
sequence saturation level.  This discrepancy is easily understood by
recalling that PMS stars around 1 Myr, as in the ONC, typically have an
order of magnitude greater surface area and hence bolometric luminosity
than main sequence stars of the same mass.}.

However, these dramatic differences do not necessarily exclude the
application of a standard dynamo (\S \ref{alpha_omega.sec}) because,
based on the limited availability of Rossby numbers for ONC stars, it
appears that ONC stars lie in the `supersaturated' regime around $\log
Ro \simeq -2$ (\S \ref{main_seq.sec}).  The slight increase in $\log
L_t$ with $\log P$ seen in the full sample (Figure \ref{Lx_rot.fig}b)
might represent the rise in X-ray emissivity from the supersaturated to
the saturated regime seen in the main sequence populations. 

One argument against an $\alpha-\Omega$ dynamo is the level of
saturation:  PMS activity shows a log-mean level of $<\log L_t/L_{bol}>
= 3.8$ which is $\sim 10$ times below the saturation level seen in main
sequence stars in the $0.5-8$ keV band.  If the same process of
magnetic field generation and eruption is involved in both classes of
stars, why should the surface activity differ by so much in a
systematic fashion?  The finding that X-ray luminosities scale
approximately with stellar area (\S \ref{size.sec}) suggests saturation
at the surface, but we can not eliminate the possibility that X-ray
luminosity instead scales with stellar volume, representing a saturation
of the internal dynamo. 

We are thus led to consider dynamos where the fields are entirely
generated and amplified in the turbulent convection zone that fills all
or most of the stellar interior (\S \ref{dist_dyn.sec}).  Such fields
may be generated both on small-scales due to turbulence in the
convection zone \citep{Durney93}, and on large scales driven by a small
differential rotation within the interior \citep[][and references
therein]{Kuker01, Kitchatinov01b}.  While a full suite of calculations is not yet
available, the solutions appear to be largely independent of the global
rotation rate, consistent with the absence of an $\log L_t-P$ relation
in our findings.  These analytical treatments are supported by recent
three-dimensional magnetohydrodynamical calculations: fields quickly
form and amplify to energy densities $>10$\% of the turbulent kinetic
energy density in both slab geometries \citep{Thelen00} and large-scale
differentially rotating spherical geometries \citep{Brun02}.  The cause
and level of saturation of these distributed dynamos are perhaps not yet 
clear\footnote{One definite prediction of distributed turbulent dynamo
models is that magnetic cycles, such as the 22-year solar oscillation,
should be absent.  Unfortunately, it will be difficult or impossible to
test this in PMS stars using X-rays as the magnetic indicator.  First,
the flare-dominated X-rays suffer much larger stochastic variability
than activity indicators arising from quiescent starspots.  Second,
stable X-ray instrumentation is rarely available for most than a
decade, and X-ray telescope allocations are usually too erratic to give
densely sampled time series over many years.}.

An important constraint on any explanation for PMS X-rays is the change
in behavior seen amoung the more massive $2-3$ M$_\odot$ stars
considered here (\S 4.3).  They exhibit an enormous dispersion in X-ray
emissivity with some in the $\log L_t/L_{bol} \simeq -4 \pm 1$ range
similar to lower mass stars, both others show $\log L_t/L_{bol} \simeq
-5 \pm 1$.  This emissivity drops further to $\log L_t/L_{bol} \sim
-8$ for B stars (F02a).  We consider two explanations for this effect,
both of which may be operative:  \begin{enumerate}

\item  Following F02a (their \S 5.2), these very low X-ray emissivities
may be misleading due to binarity, where a lower mass secondary
produces the observed X-rays and the higher mass p[rimary (which
dominates $L_{bol}$) is magnetically inactive.  The X-ray luminosities
of these systems are somewhat higher than the average low-mass PMS ONC
stars, implying that the companions have higher than average mass (e.g.
1 M$_\odot$ rather than 0.3 M$_\odot$).  Detailed optical study of the
$2-3$ M$_\odot$ population could test the binarity hypothesis.

\item The drop in X-ray emissivity among intermediate mass PMS stars by
an order of magnitude (or more if the binary hypothesis is correct) may
be linked to structural changes in the stellar interior and consequent
changes in dynamo activity.  \citet{Palla93} show that PMS stars with
masses above $\simeq 4$ M$_\odot$ arrive at the stellar birthline with
radiative interiors undergoing nonhomologous contraction, in contrast
to PMS stars below $M \simeq 2$ M$_\odot$ with fully convective
interiors undergoing homologous contraction.  They predict a narrow
range of PMS masses, $2.4 < M < 3.9$ M$_\odot$ in their canonical
model, where a composite structure of radiative core and convective
mantle heated by deuterium burning occurs.  The precise boundaries of
these structural changes are very sensitive to the initial conditions,
so that intermediate-mass ONC stars with somewhat different ages and
accretion histories can have very different structures.  These internal
structure differences may be reflected in the efficiency of the
magnetic dynamo, leading to the wide dispersion of $\log L_t/L_{bol}$
ratios we see in the $2 < M < 3$ M$_\odot$ mass range (Figure
\ref{Lx_M.fig}c).  The exact nature of the magnetic fields in these
stars is not clear:  conceivably different combinations of a
distributed dynamo, tachocline dynamo or fossil field could be present
in different stars with similar masses.

\end{enumerate}

\subsection{Implications for other models}

While models of relic and core magnetic fields in PMS stars are not
fully developed (\S \ref{relic.sec}), our findings do not support these
as the source of fields responsible for the observed X-ray emission.
We find only a mild temporal dependence of X-ray luminosity on stellar
age ranging from $10^5$ to $10^7$ yr (\S \ref{age.sec}), during which
time the stellar interior undergoes the important transition to a
radiative core.  The only hint of a dependence on internal structure is
the clear dependence of X-rays on stellar mass.  However, we cannot
determine whether the $L_t-M$ relationship arises from an astrophysical
mechanism or as a byproduct of more fundamental relationships like
$L_t/L_{bol} \propto $ constant, $L_t \propto R^2$, or $L_{bol} \propto
M$.  But if a causal link between magnetic activity and mass is
present, it conceivably could arise from the increased trapping of
relic fields during the gravitational collapse of more massive stars,
or from the increased capture of flux in the radiative core of more
massive stars.

Our results also lend little support for models where X-ray emission is
associated with a circumstellar disk (\S
\ref{disk_fields.sec})\footnote{After this paper was submitted, a
closely related study by \citet{Flaccomio02} was released.  It is based
on the detection of 342 out of 696 unabsorbed ONC stars with the
$Chandra$ High Resolution Camera which, unlike ACIS, does not give
spectral information on the sources.  Many of their findings are
similar to ours: no correlation of $\log L_x/L_{bol}$ with rotational
period, strong correlation between $\log L_x$ and mass with
low-luminosity outliers at intermediate masses; and a decline of $\log
L_x$ with stellar age.  One difference concerns the X-ray relationship
to circumstellar disks: using the strength of the Ca II triplet lines
as an indicator of accretion (in contrast to our photometric and
imaging disk indicators), they find that low-accretion ONC stars have
an order of magnitude higher X-ray luminosity than high-accretion
stars.  This is further evidence that young stellar X-rays are not
primarily produced in disk fields.}.  This result may have important
implications for the physics of the circumstellar disk; in particular,
X-ray ionization of disk gas and energetic particle bombardment of disk
solids should be lower if the X-rays arise from fields close to the
stellar surface than if they arise from the immediate vicinity of the
disk \citep[][F02b]{Glassgold00}.

Finally, we note that the activity-rotation diagram for PMS stars bears
some phenomenological similarity to that obtained for post-main
sequence giants and main sequence dM stars (\S \ref{dM_giant.sec}).
For example, intermediate-mass $2-3$ M$_\odot$ giants and PMS stars
show the same wide range of X-ray luminosities, from $\log L_x < 28$
erg s$^{-1}$ to 31 erg s$^{-1}$, unaffected by a wide range of
rotational velocities \citep{Pizzolato00}. dM stars show a strong link
between $\log L_x$ and stellar size \citep{Houdebine97}.  However, as
several different models still compete to explain activity in these stars,
it is unclear whether phenemological similarities between the magnetic
activity of PMS, dM and giant stars are astrophysically meaningful.

\section{Concluding comments \label{concl.sec}}

With the greatly enlarged sample provided by the ONC, observational
constraints on the origins of magnetic activity in low-mass PMS stars
are more quantitatively and securely established compared to previous
results.  But, at present, we can not establish a definitive link
between our findings and a unique theory of magnetic field generation
in PMS stars.

There are two sources of uncertainty.  First, we encounter a degeneracy
between the physical properties correlated with X-ray emission.
Examination of the evolutionary tracks in the HR diagram immediately
reveals that bolometric luminosity, radius, mass and age are mutually
dependent in a non-trivial and systematic fashion.  We thus can not
confidently extract from observations alone which property is
astrophysically responsible for the magnetic activity we detect.

Second, theoretical models have often not been sufficiently developed
to compare with our empirical findings; additional theoretical
calculations are clearly necessary. For example, calculation of Rossby
numbers \citep[as in][]{Kim96} for each star in our sample using its
specific mass, age and rotation, would populate the X-ray/Rossby number
diagram (Figure \ref{Lx_Rossby.fig}) and possibly reveal new
constraints and trends.  It would also be very useful if PMS dynamo
models involving $\alpha-\Omega$, $\alpha-\alpha$ and other distributed
field generation processes were produced for PMS interiors with a wide
range of masses and rotations for comparison with our findings.
Initial models of this type have been reported by
\citet{Kitchatinov01b, Kuker01} and references therein.

Despite these difficulties, the results seem to favor certain
interpretations.  The absence of an activity-rotation relation is by
itself a good argument for some form of distributed dynamo arising
throughout the convective zone, rather than the standard
$\alpha-\Omega$ dynamo involving a tachocline.  The scaling between
X-ray emission and the volume of the convective region at lower masses,
and the change of X-ray properties in some stars at intermediate masses
when a radiative core appears, together support a distributed dynamo
for most T Tauri stars.

But we cannot yet exclude alternatives such as a standard dynamo in a
`saturated' or `supersaturated' regime, where the saturation level
occurs at a substantially lower value of $L_x/L_{bol}$ than in main
sequence stars.  If PMS stars indeed all have `saturated' dynamos, it
is possible that little will be learned of their magnetic processes,
expecially as we do not understand the causes of saturation even in
main sequence stars. Similarly, magnetic reconnection of a
mass-dependent fossil field may still be a viable model.  However, the
findings to not support models where the X-rays are associated with a
circumstellar disk, either reconnection of star-disk fields at the
corotation radius or reconnection of sheared disk-disk fields.

Additional forthcoming X-ray observations of the ONC should provide
critical new insights.  A contiguous $Chandra$ ACIS observation
spanning $\simeq 11$ days is planned which will give an order of
magnitude increase in sensitivity essential for tracing magnetic
activity in $M \leq 0.7$ M$_\odot$ PMS stars, and a sufficiently long
time series of all stars to obtain detailed characteristics of PMS
X-ray emission.  Several relevant studies are planned.  The statistical
properties of X-ray flares (e.g., the distribution of energies,
durations and recurrence rates) may reveal similarities or differences
when compared to flares in the Sun and older active stars.  Quiescent
X-ray levels between flares will be sought, and may show less scatter
in correlations with other stellar properties than we find here.  We
will search for rotationally modulated X-ray emitting structures which
might reveal large-scale asymmetries in the magnetic field geometry
predicted by $\alpha-\alpha$ dynamos and relic core fields.
Conceivably, transitions in the levels and structure of surface
magnetic fields reflecting the emergence of a core radiative zone will
be seen in comparisons of younger $vs.$ older and less $vs.$ more
massive PMS stars.

\acknowledgements{We are greatly appreciative of the careful and
insightful reading of the manuscript by Dermott Mullan (Bartol) and the
anonymous referee.  EDF also greatly benefited from discussions with
participants of stellar magnetism workshops in Santiago, Boulder and
Toulouse during 2001$-$02.  Patrick Broos (Penn State), Steven Pravdo
(JPL), and Yohko Tsuboi (Penn State/Chuo) played critical roles in the
$Chandra$ ACIS Orion project.  Sofia Randich (Arcetri) provided
valuable help and comments.  This work was principally supported by
NASA contract NAS 8-38252 (Garmire, PI).}

\newpage

\clearpage
\newpage
\begin{figure}
\centering
  \begin{minipage}[t]{1.0\textwidth}
  \centering
  \includegraphics[height=0.45\textheight]{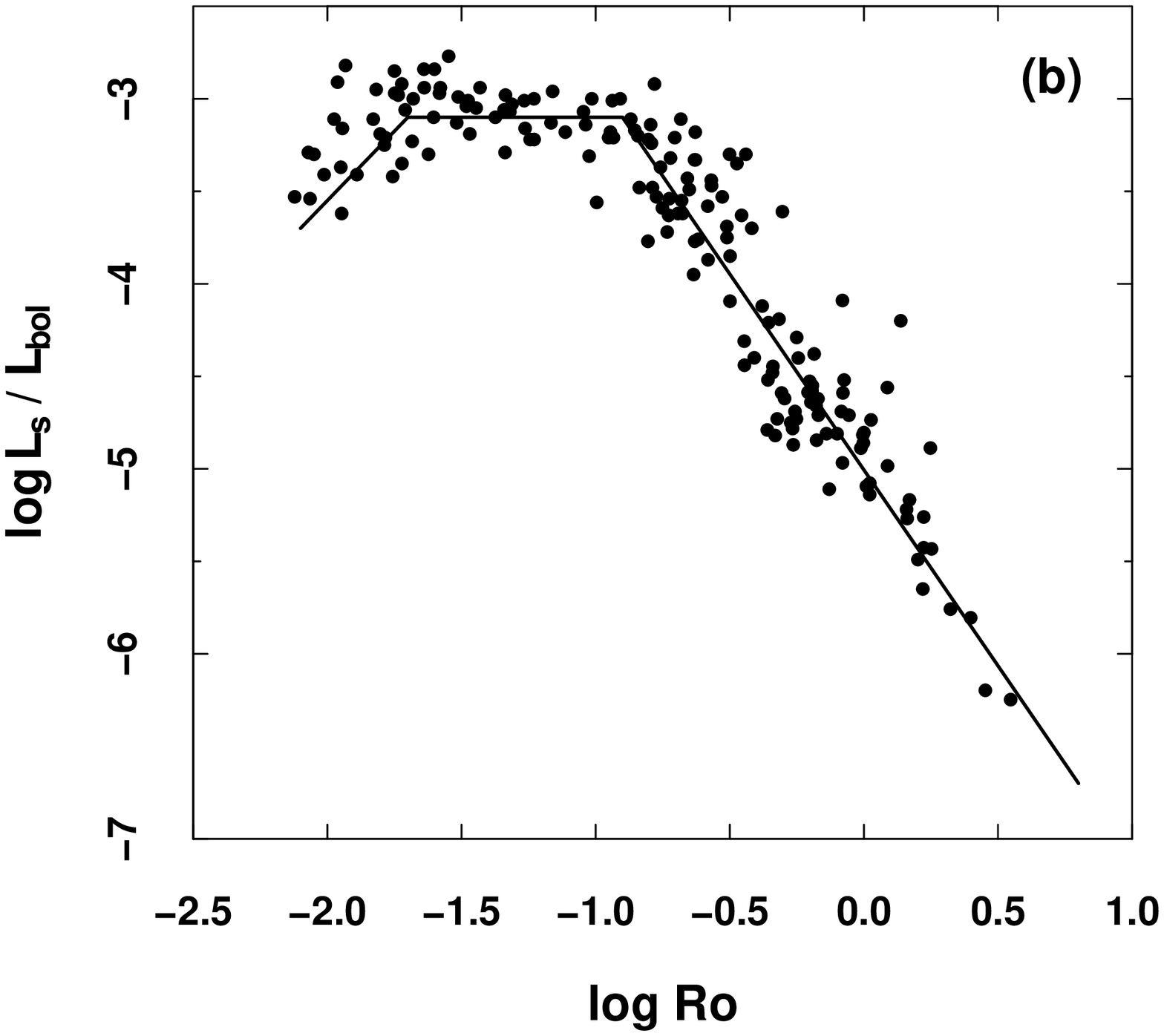}
  \end{minipage} \\ [0.0in]
  \begin{minipage}[t]{1.0\textwidth}
  \centering
  \includegraphics[height=0.45\textheight]{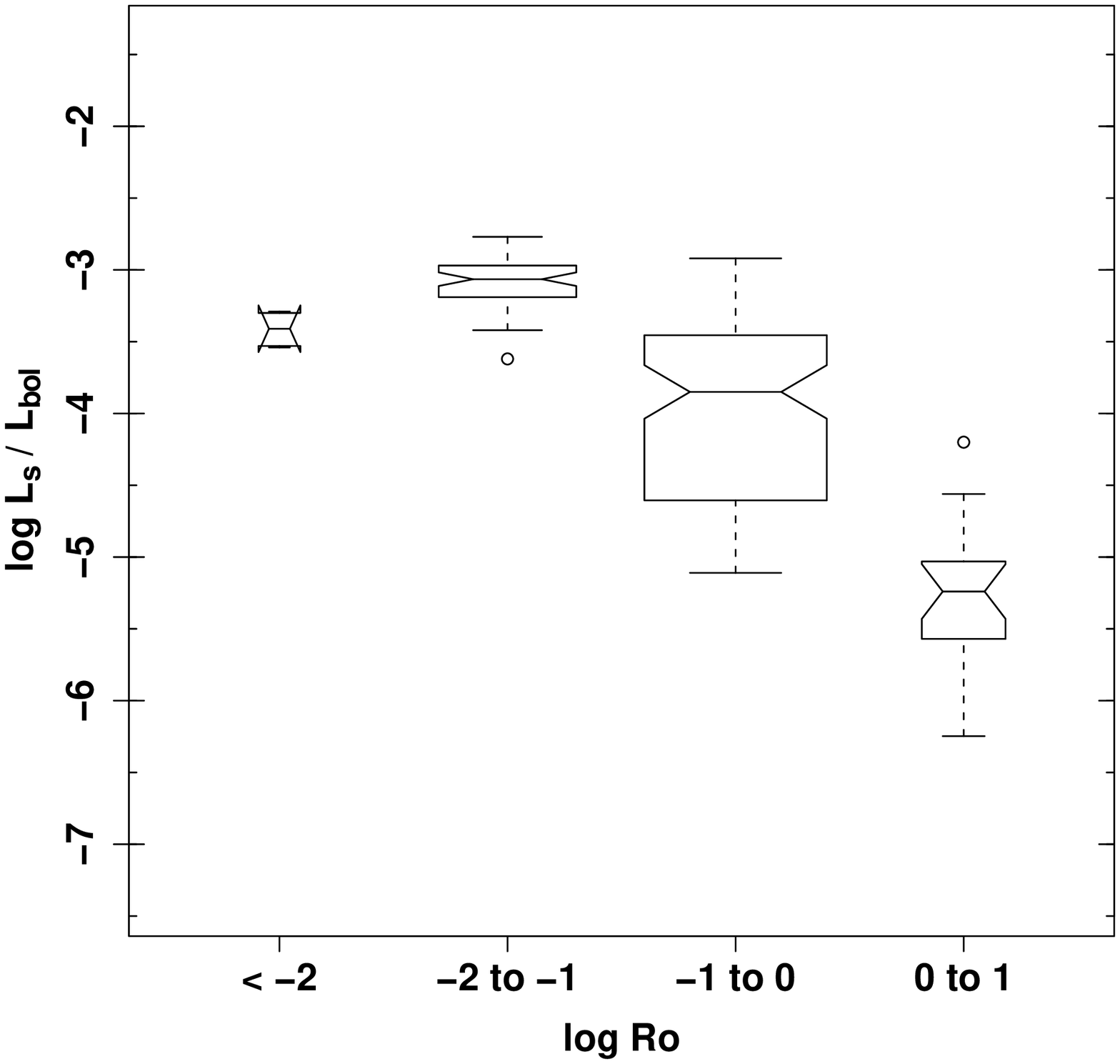}
  \end{minipage}
\end{figure}

\clearpage
\newpage

\begin{figure}
\centering
 \begin{minipage}[t]{1.0\textwidth}
  \centering
  \includegraphics[height=0.45\textheight]{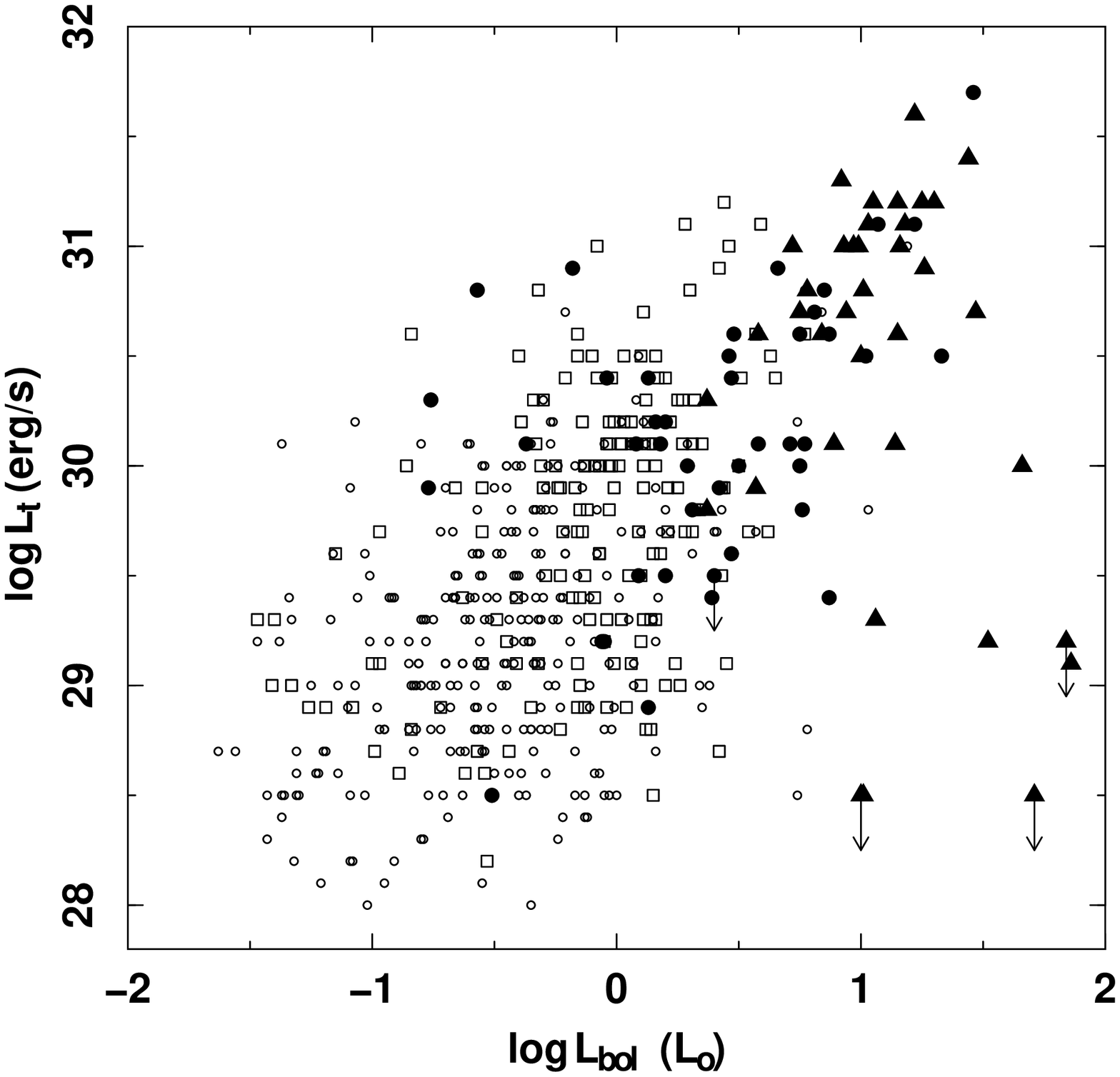}
  \end{minipage} \\ [0.0in]
  \begin{minipage}[t]{1.0\textwidth}
  \centering
  \includegraphics[height=0.45\textheight]{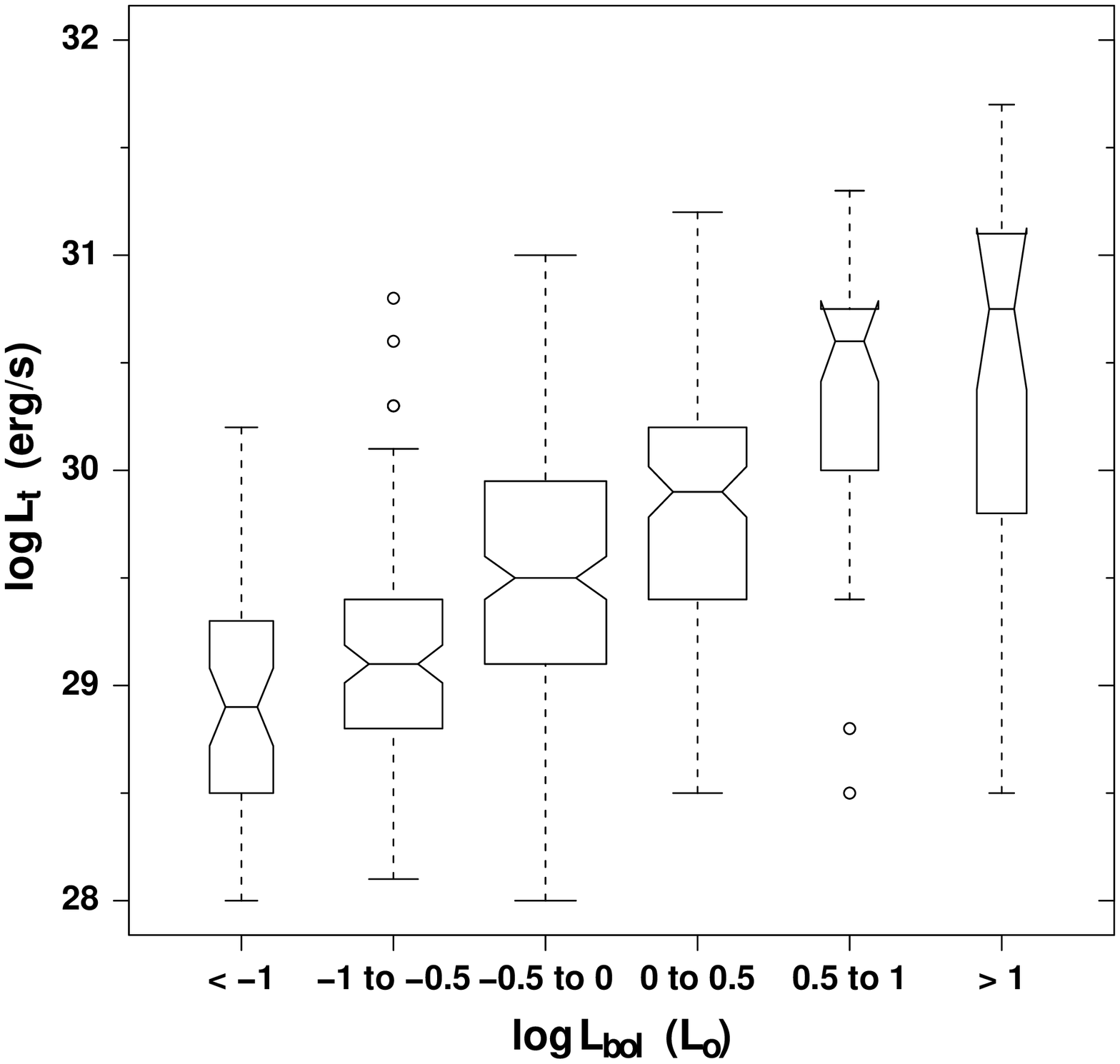}
  \end{minipage}
\caption{Relationship between PMS X-rays and bolometric luminosities:
(a) scatter plot of $\log L_t$ and $\log L_{bol}$, and (b) boxplot of
$\log L_t$ and $\log L_{bol}$.  The scatter plot symbols here and in
later figures are coded by mass as follows:  $1.4 < M < 3.0$ M$_\odot$
(large filled triangles); $0.7 < M < 1.4$ M$_\odot$ (large filled
circles); $0.25 < M < 0.7$ M$_\odot$ (open squares); and $M < 0.25$
M$_\odot$ (small open circles).  The four X-ray non-detections with $M
> 0.7$ M$_\odot$ are shown with arrows.  
\label{Lx_Lbol.fig}}
\end{figure}

\clearpage
\newpage

\begin{figure}
\centering
 \begin{minipage}[t]{1.0\textwidth}
  \centering
  \includegraphics[height=0.45\textheight]{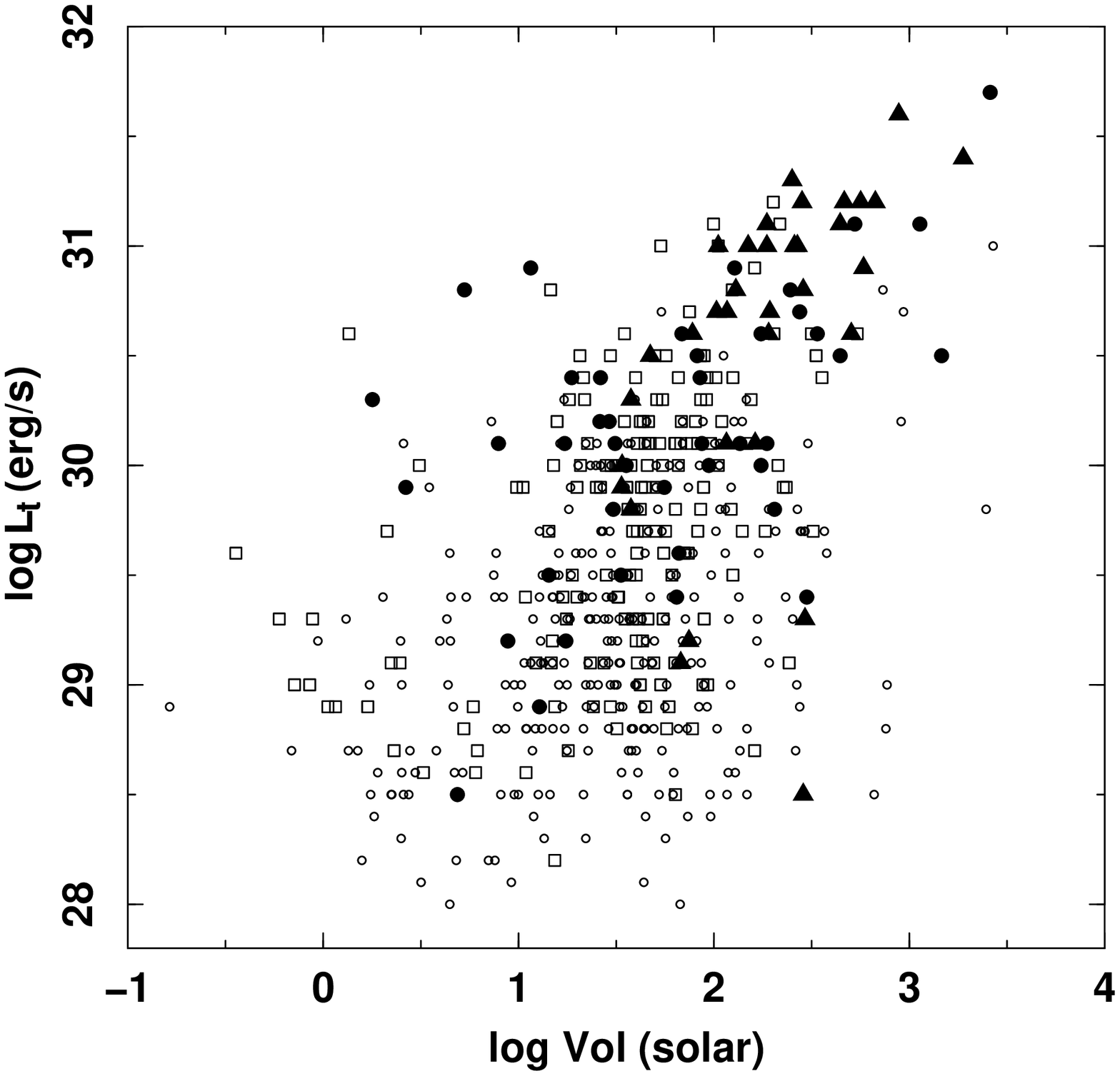}
  \end{minipage} \\ [0.0in]
  \begin{minipage}[t]{1.0\textwidth}
  \centering
  \includegraphics[height=0.45\textheight]{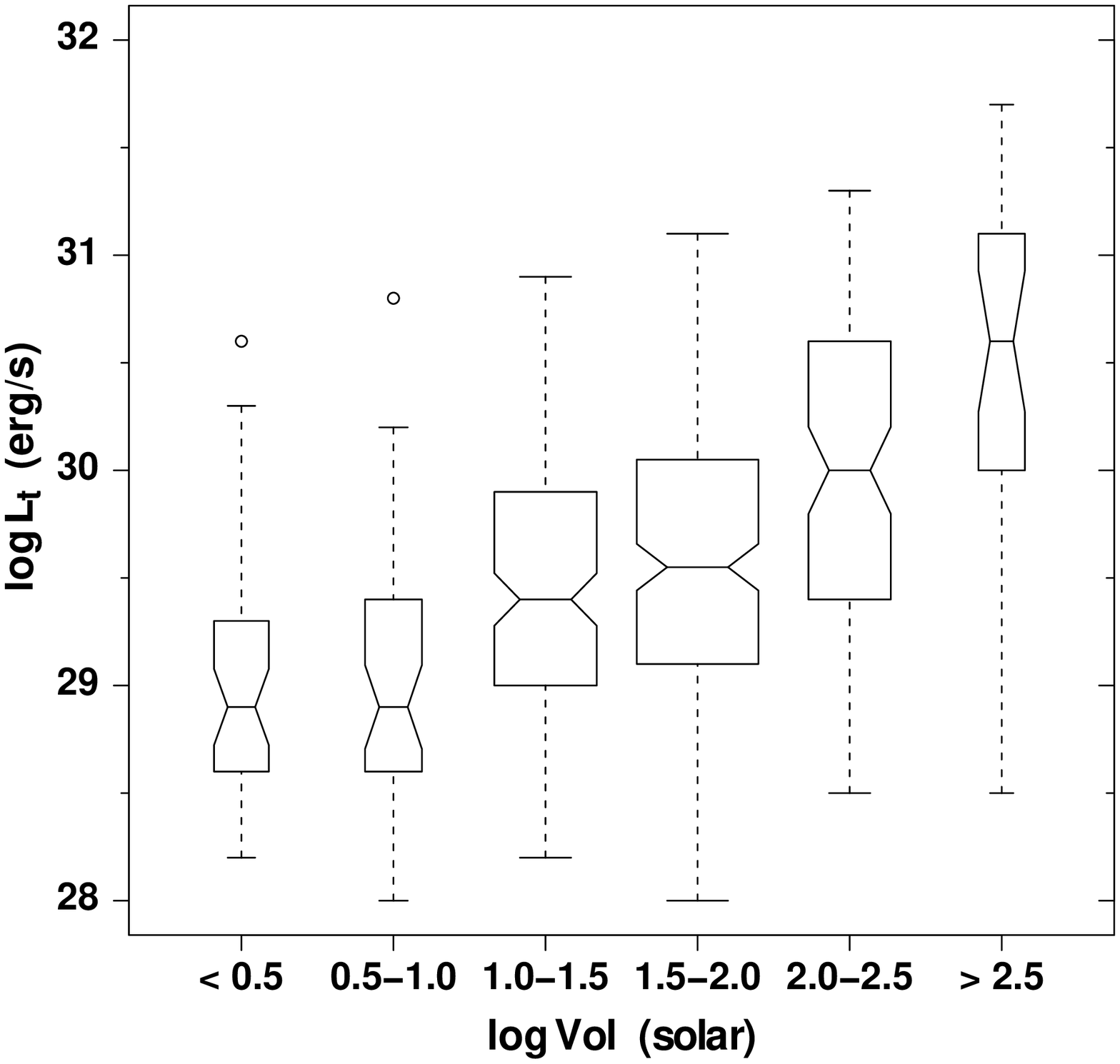}
  \end{minipage}
\caption{Relationship between PMS X-rays and stellar volumes:  (a)
scatter plot of $\log L_t$ and $\log V$, (b) boxplot of $\log L_t$ and
$\log V$, (c) scatter plot of $\log L_t/L_{bol}$ and $\log V$, (d)
boxplot of $\log L_t/L_{bol}$ and $\log V$.  See Figure
\ref{Lx_Lbol.fig} caption for symbol definitions.
\label{Lx_V.fig}}
\end{figure}

\clearpage
\newpage

\begin{figure}
 \begin{minipage}[t]{1.0\textwidth}
  \centering
  \includegraphics[height=0.45\textheight]{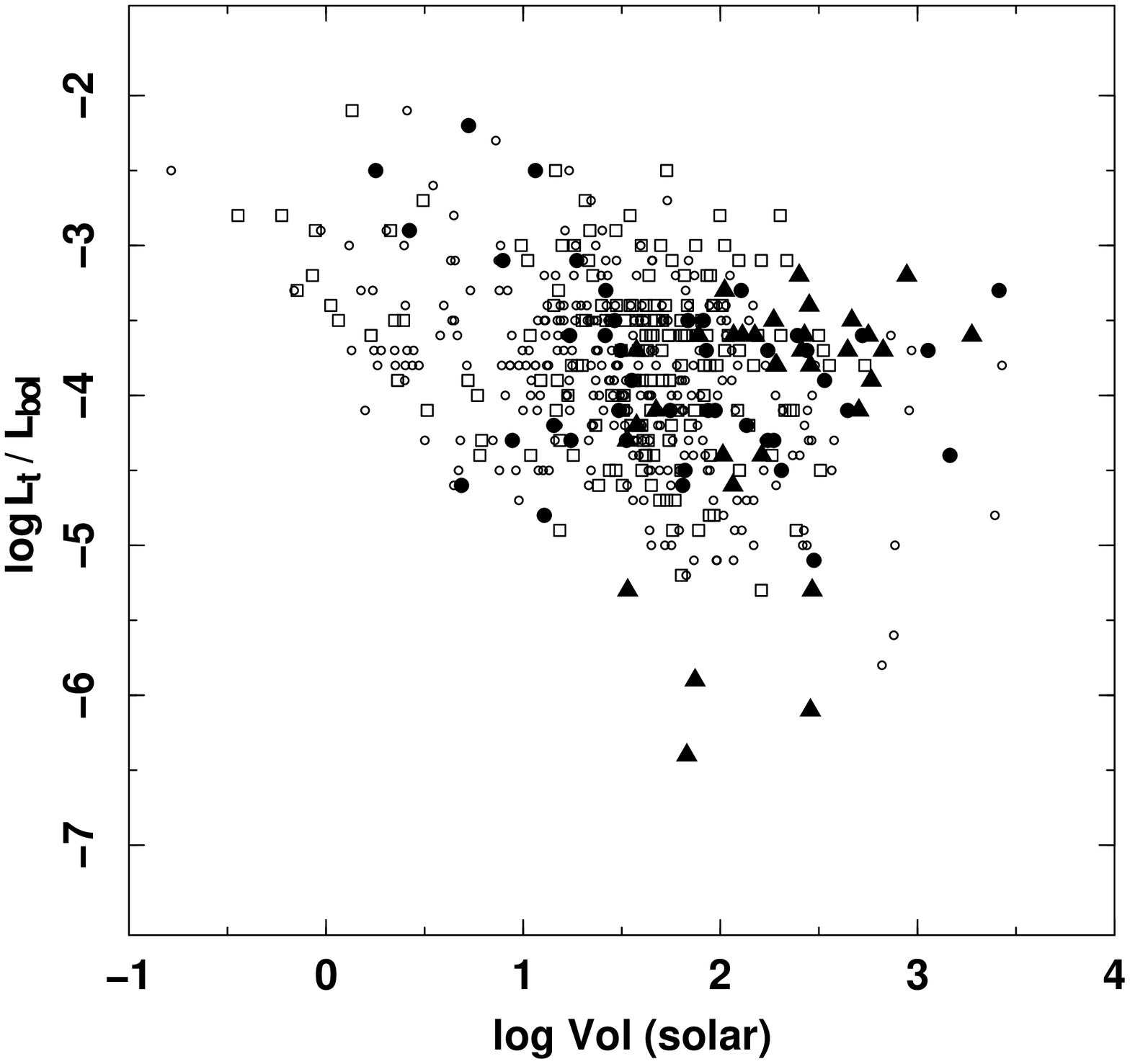}
  \end{minipage} \\ [0.0in]
  \begin{minipage}[t]{1.0\textwidth}
  \centering
  \includegraphics[height=0.45\textheight]{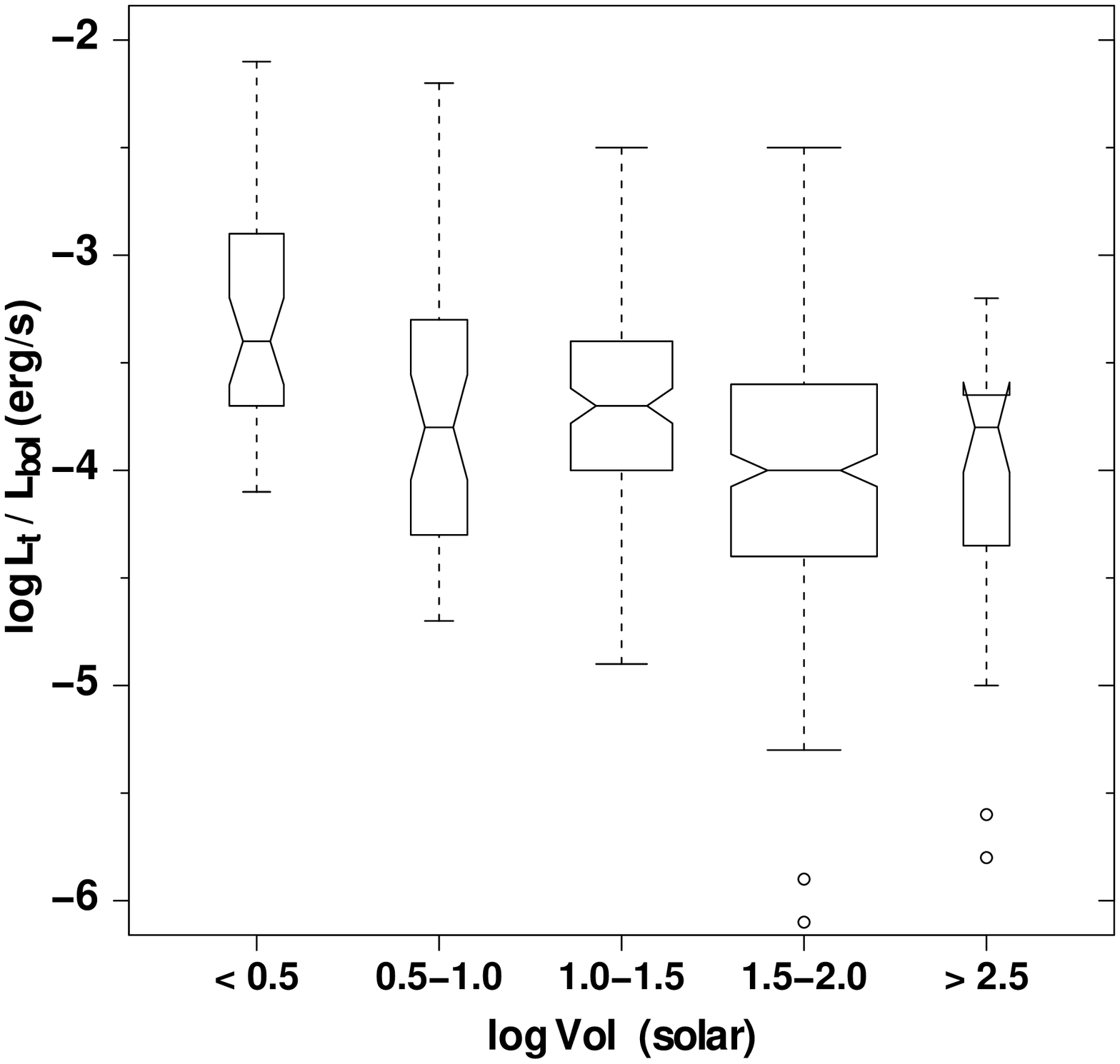}
  \end{minipage}
\end{figure}

\clearpage
\newpage

\begin{figure}
 \begin{minipage}[t]{1.0\textwidth}
  \centering
  \includegraphics[height=0.45\textheight]{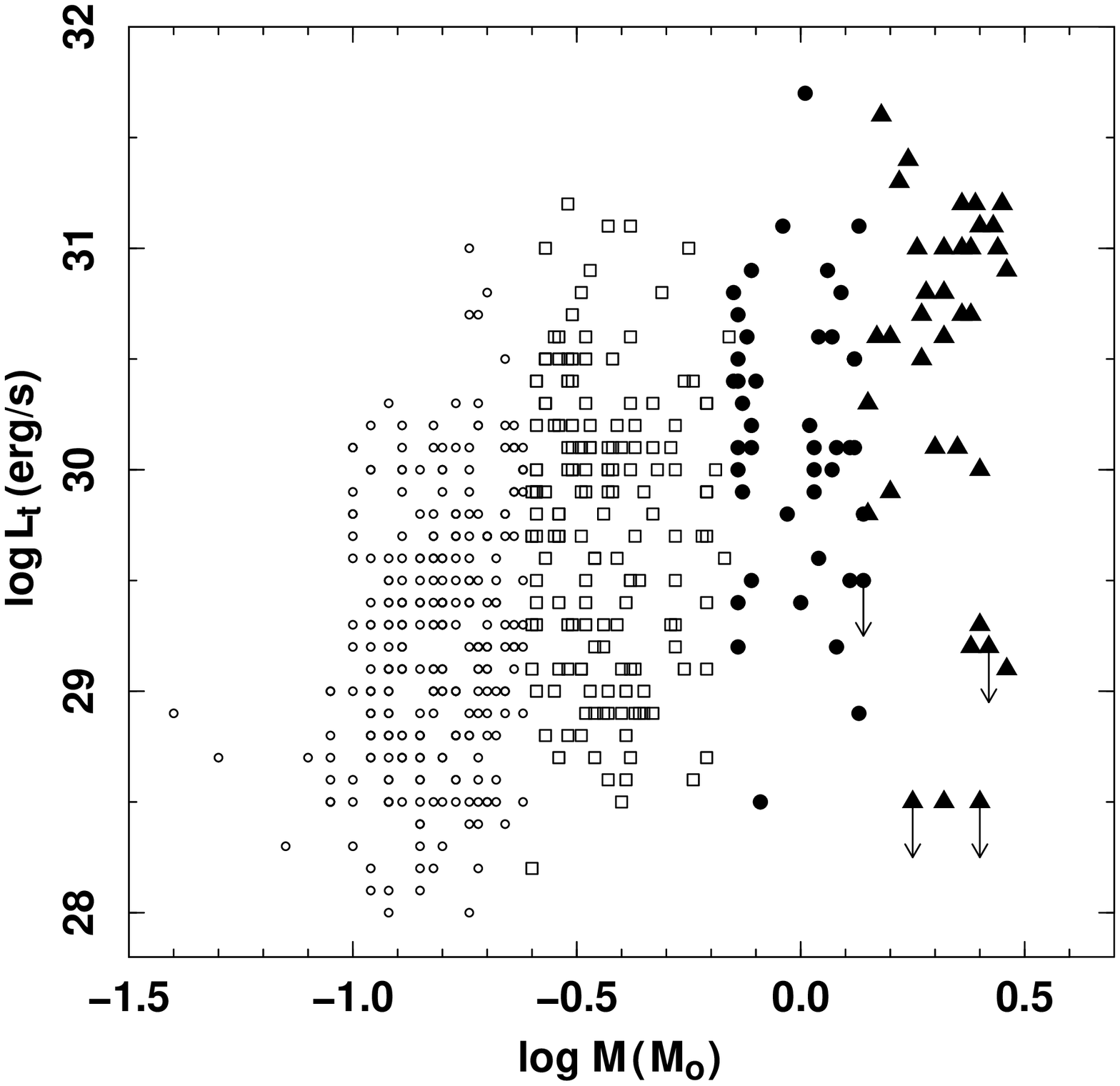}
  \end{minipage} \\ [0.0in]
  \begin{minipage}[t]{1.0\textwidth}
  \centering
  \includegraphics[height=0.45\textheight]{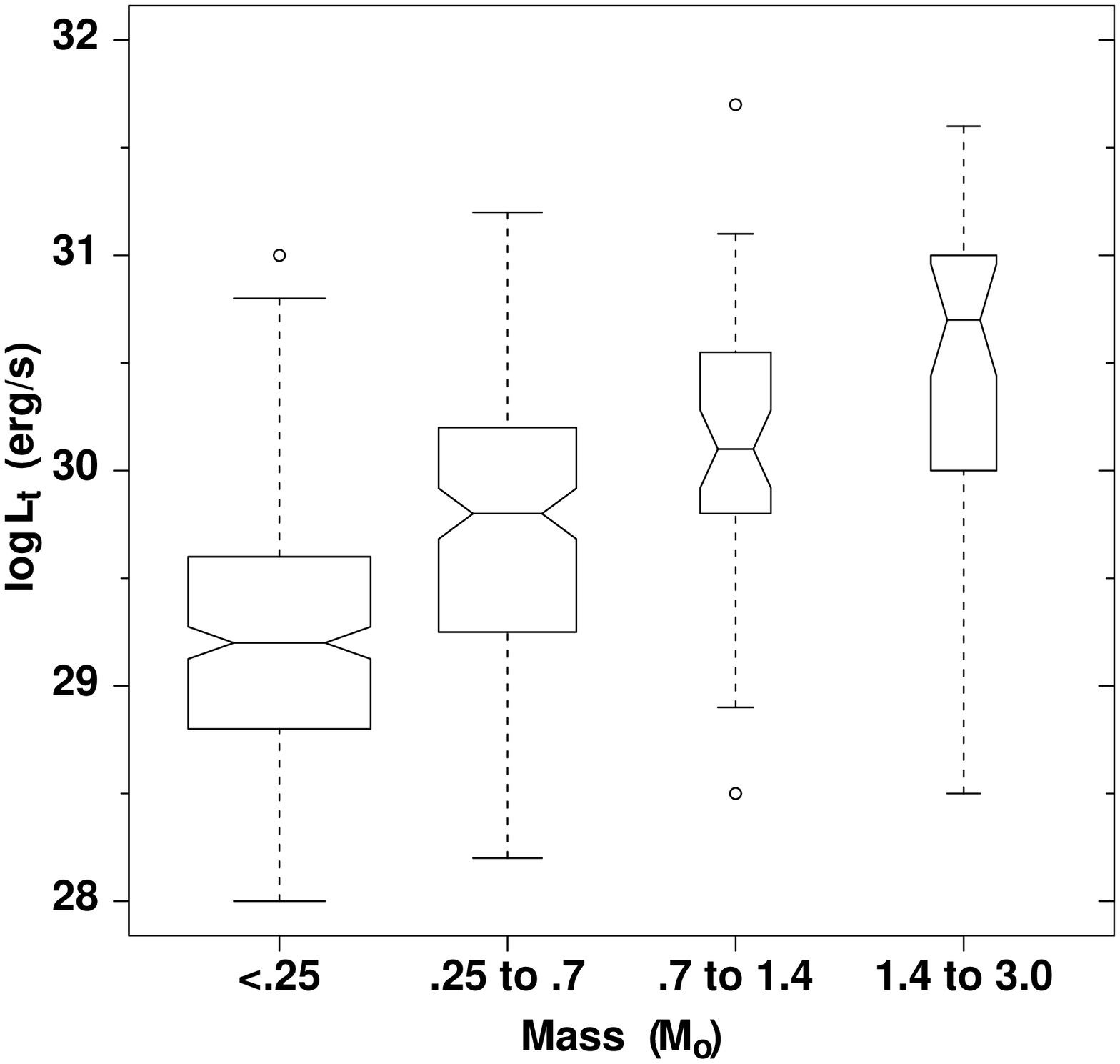}
  \end{minipage}
\caption{Relationship between PMS X-rays and stellar masses:  (a)
scatter plot of $\log L_t$ and $\log M$, (b) boxplot of $\log L_t$ and
$\log M$, (c) scatter plot of $\log L_t/L_{bol}$ and $\log M$, (d)
boxplot of $\log L_t/L_{bol}$ and $\log M$.  See Figure
\ref{Lx_Lbol.fig} caption for symbol definitions.
\label{Lx_M.fig}}
\end{figure}

\clearpage
\newpage

\begin{figure}
 \begin{minipage}[t]{1.0\textwidth}
  \centering
  \includegraphics[height=0.45\textheight]{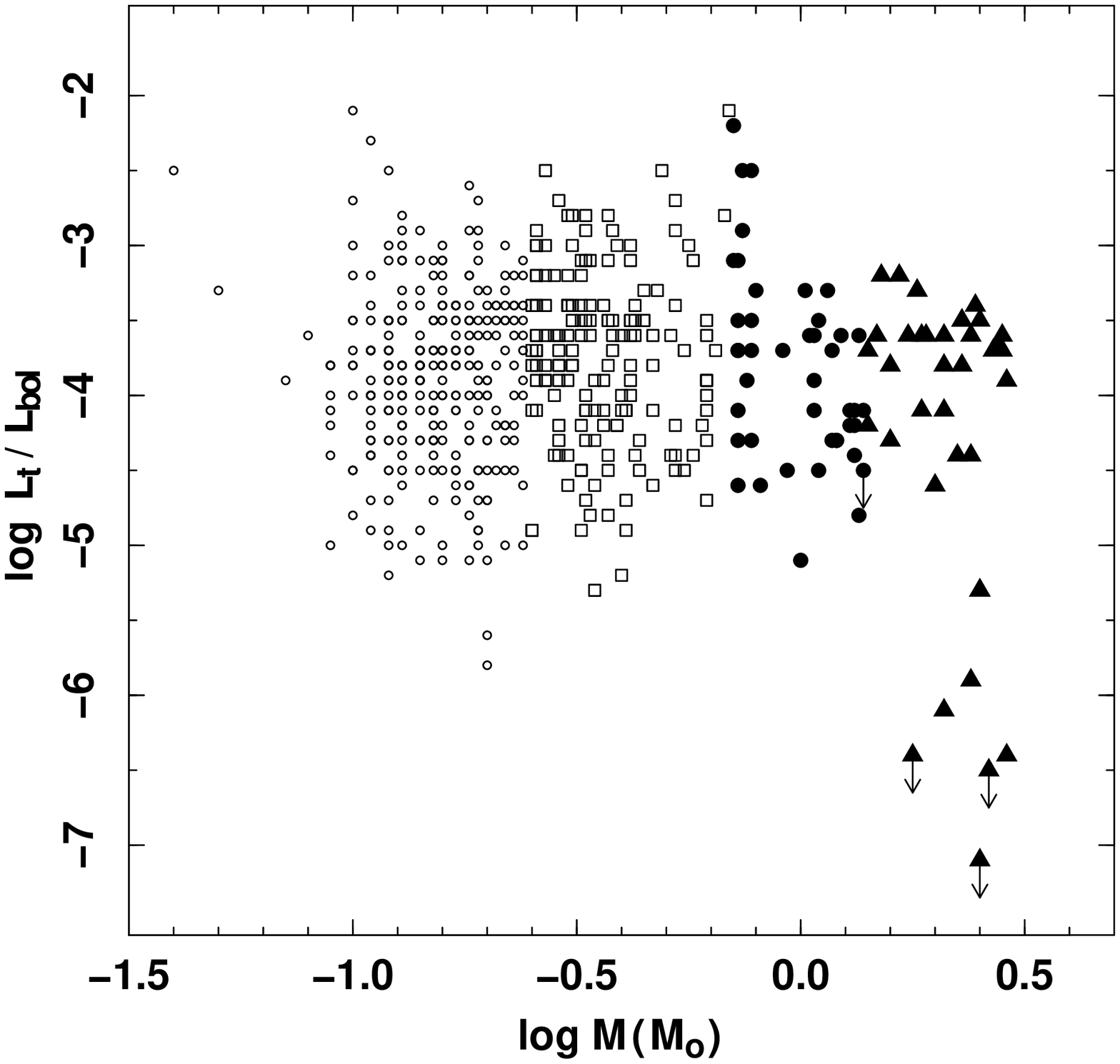}
  \end{minipage} \\ [0.0in]
  \begin{minipage}[t]{1.0\textwidth}
  \centering
  \includegraphics[height=0.45\textheight]{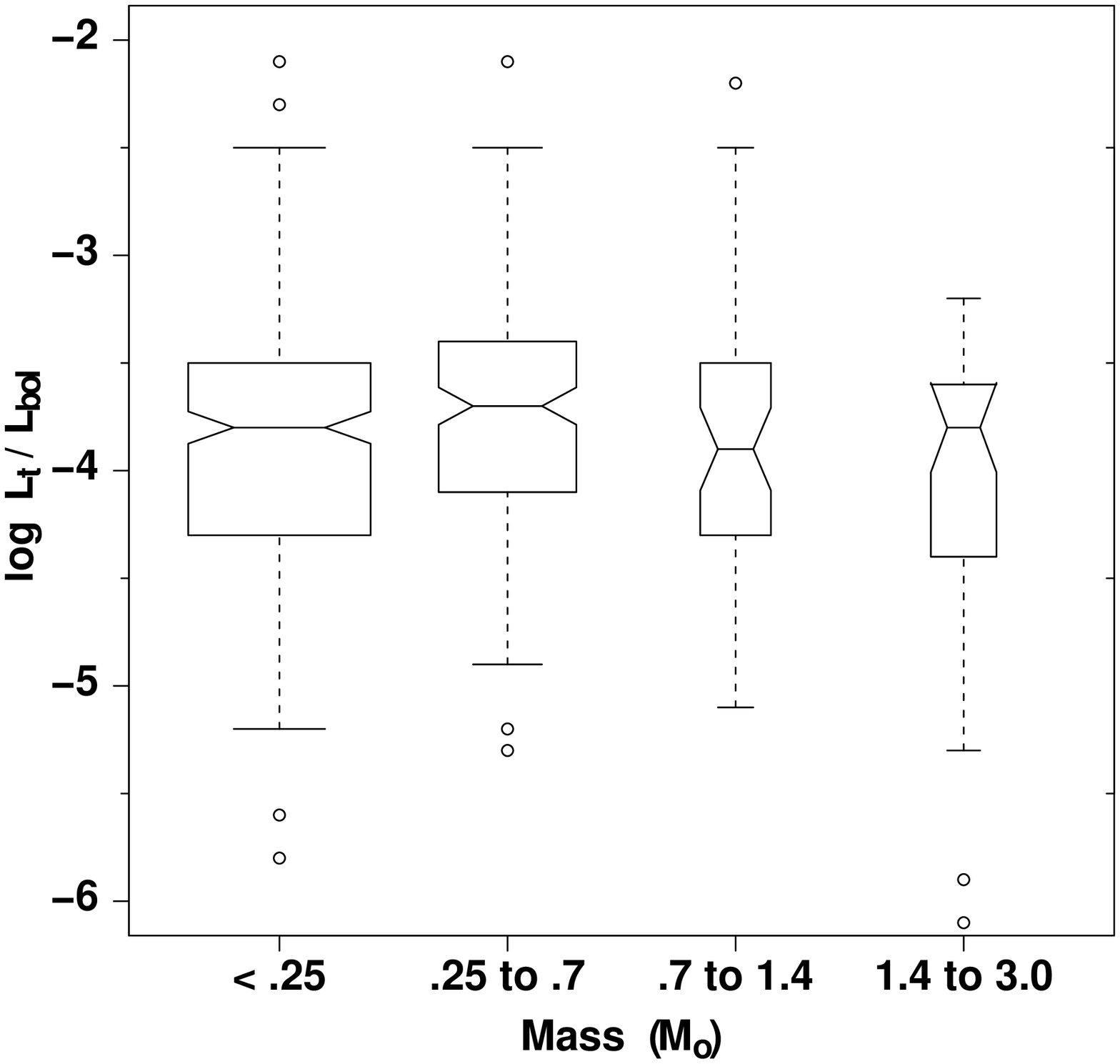}
  \end{minipage}
\end{figure}

\clearpage
\newpage

\begin{figure}
 \begin{minipage}[t]{1.0\textwidth}
  \centering
  \includegraphics[height=0.45\textheight]{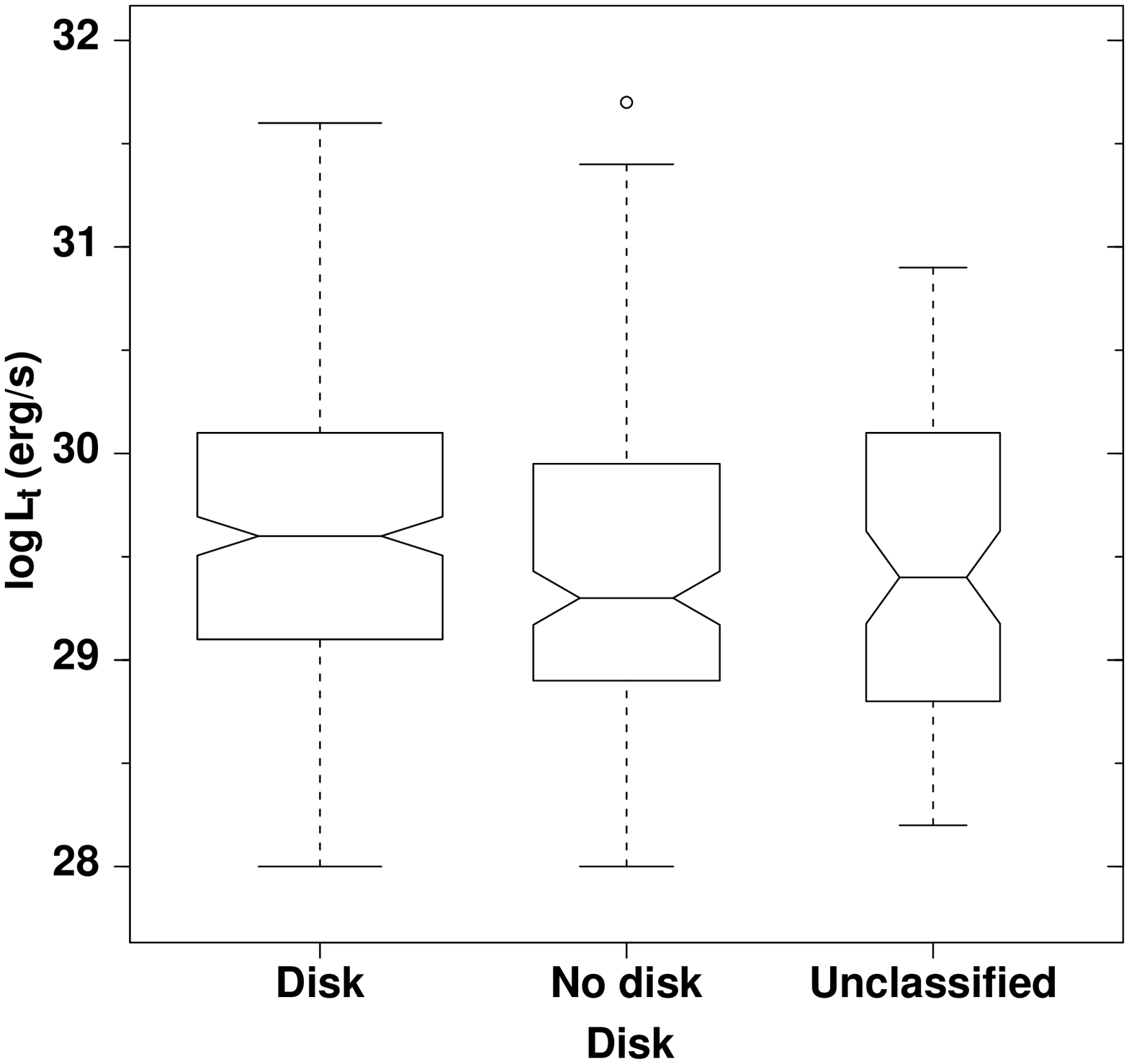}
  \end{minipage} \\ [0.0in]
  \begin{minipage}[t]{1.0\textwidth}
  \centering
  \includegraphics[height=0.45\textheight]{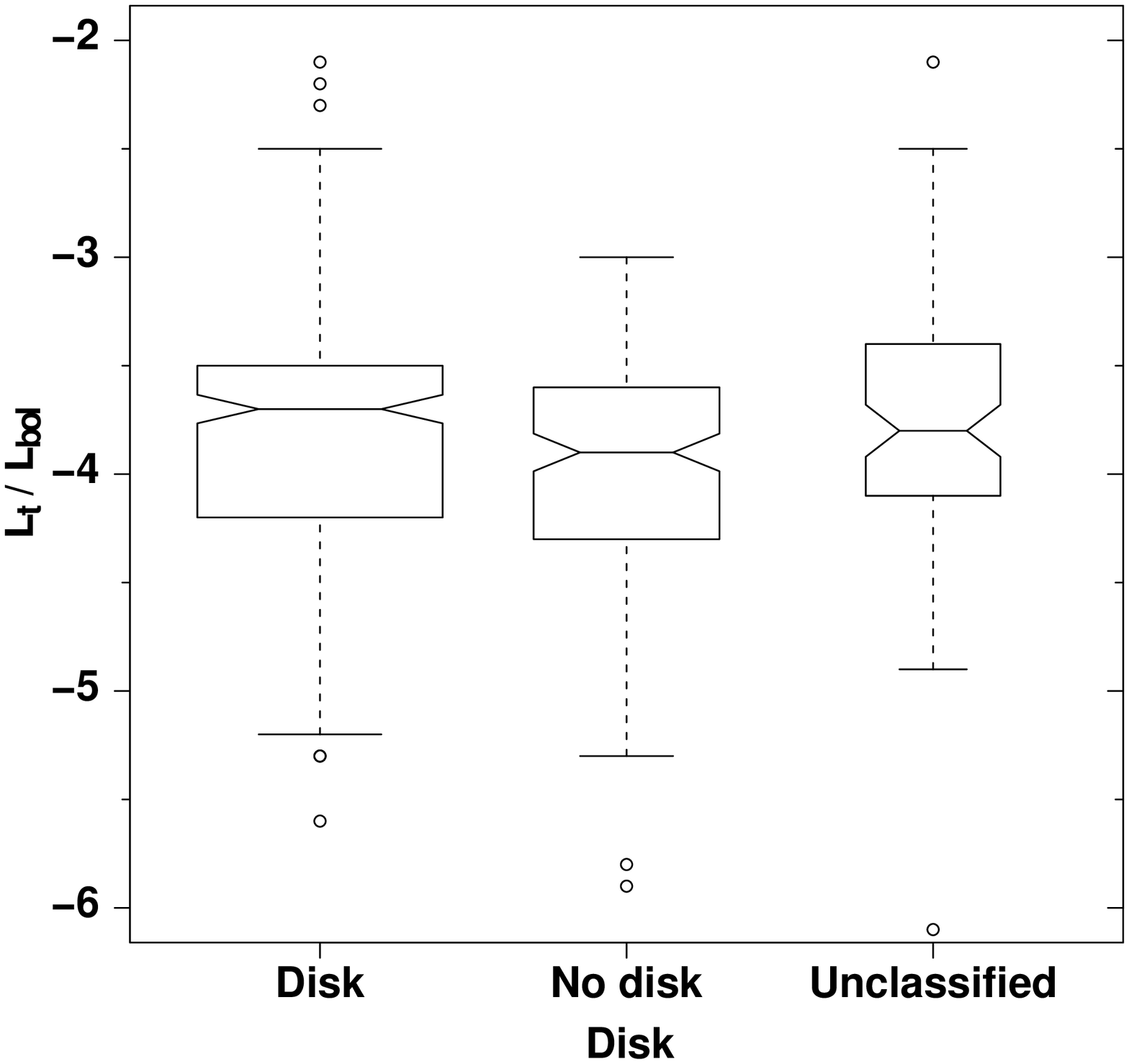}
  \end{minipage}
\caption{Relationship between PMS X-rays and circumstellar disks:
boxplots of (a) $\log L_t$, (b) $\log L_t/L_{bol}$ and (c) mass $vs.$
disk indicator.  See Figure \ref{Lx_Lbol.fig} caption for symbol
definitions.  \label{Lx_disk.fig}}
\end{figure}

\clearpage
\newpage

\begin{figure}
  \centering
  \includegraphics[height=0.45\textheight]{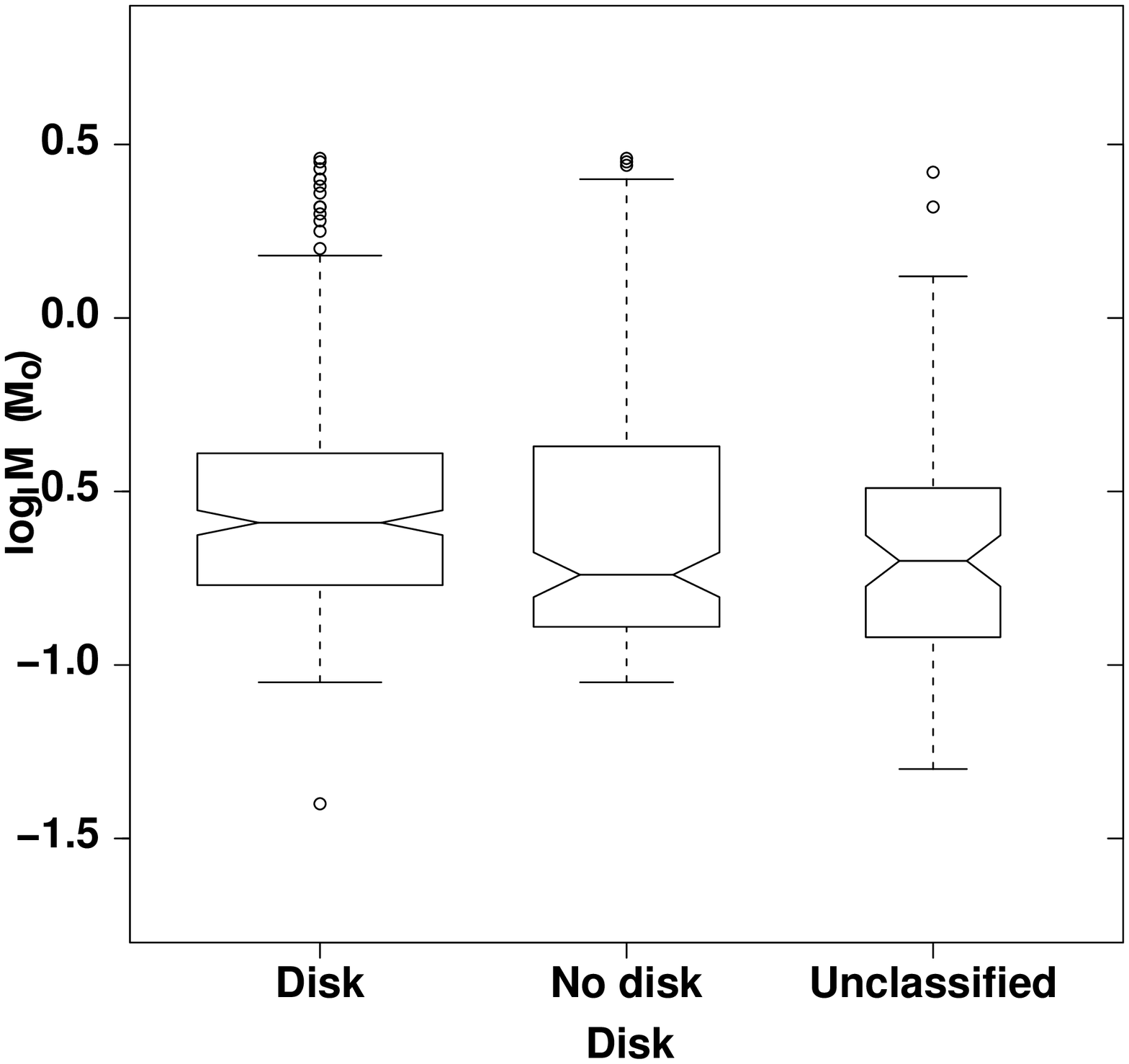}
\end{figure}

\clearpage
\newpage

\begin{figure}
\centering
 \begin{minipage}[t]{1.0\textwidth}
  \centering
  \includegraphics[height=0.45\textheight]{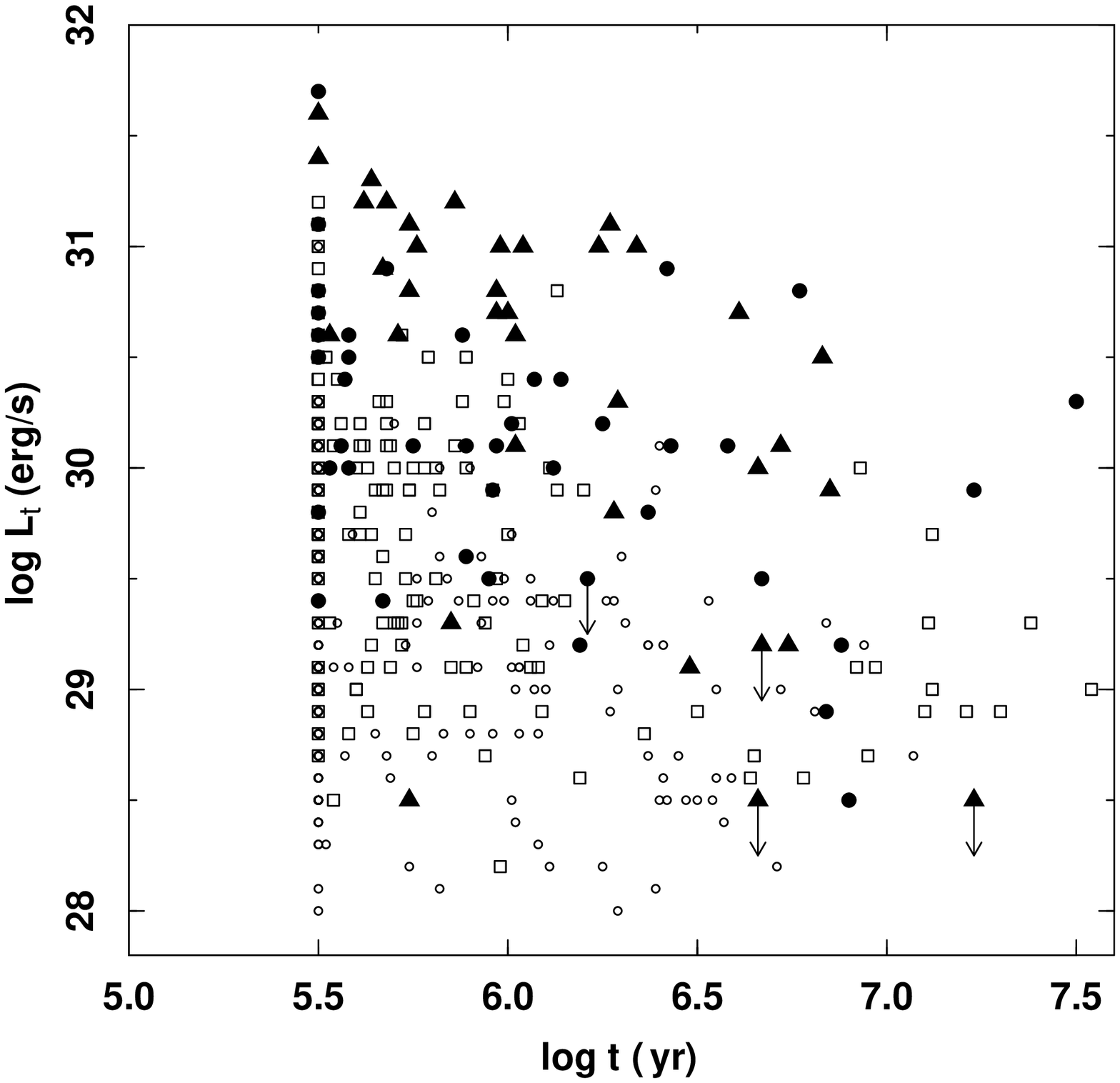}
  \end{minipage} \\ [0.0in]
  \begin{minipage}[t]{1.0\textwidth}
  \centering
  \includegraphics[height=0.45\textheight]{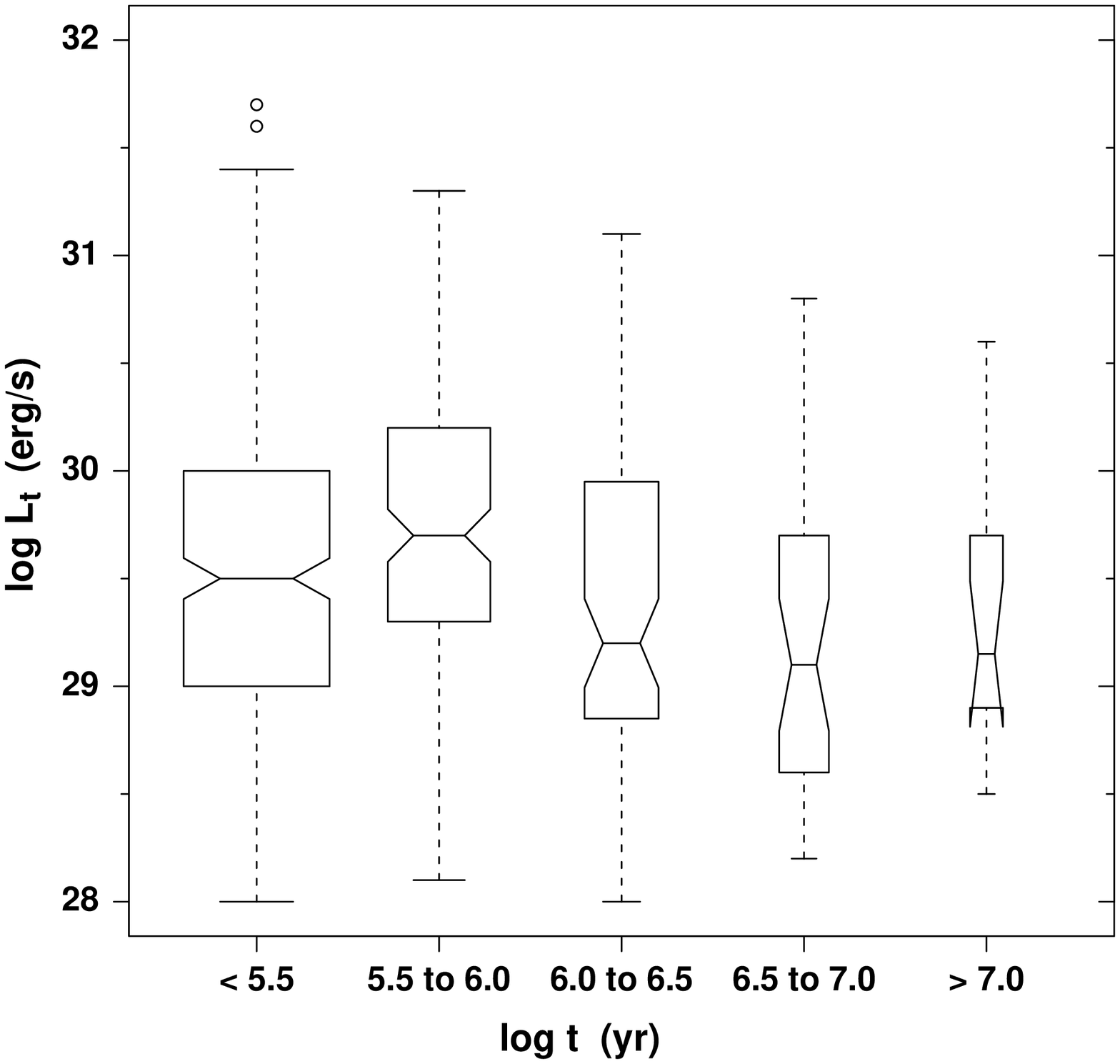}
  \end{minipage}
\caption{Relationship between PMS X-rays and stellar ages:  (a) scatter
plot of $\log L_t$ and $\log t$, (b) boxplot of $\log L_t$ and $\log
t$, (c) scatter plot of $\log L_t/L_{bol}$ and $\log t$, (d) boxplot of
$\log L_t/L_{bol}$ and $\log t$.  See Figure \ref{Lx_Lbol.fig} caption
for symbol definitions.
\label{Lx_age.fig}}
\end{figure}

\clearpage
\newpage

\begin{figure}
\centering
 \begin{minipage}[t]{1.0\textwidth}
  \centering
  \includegraphics[height=0.45\textheight]{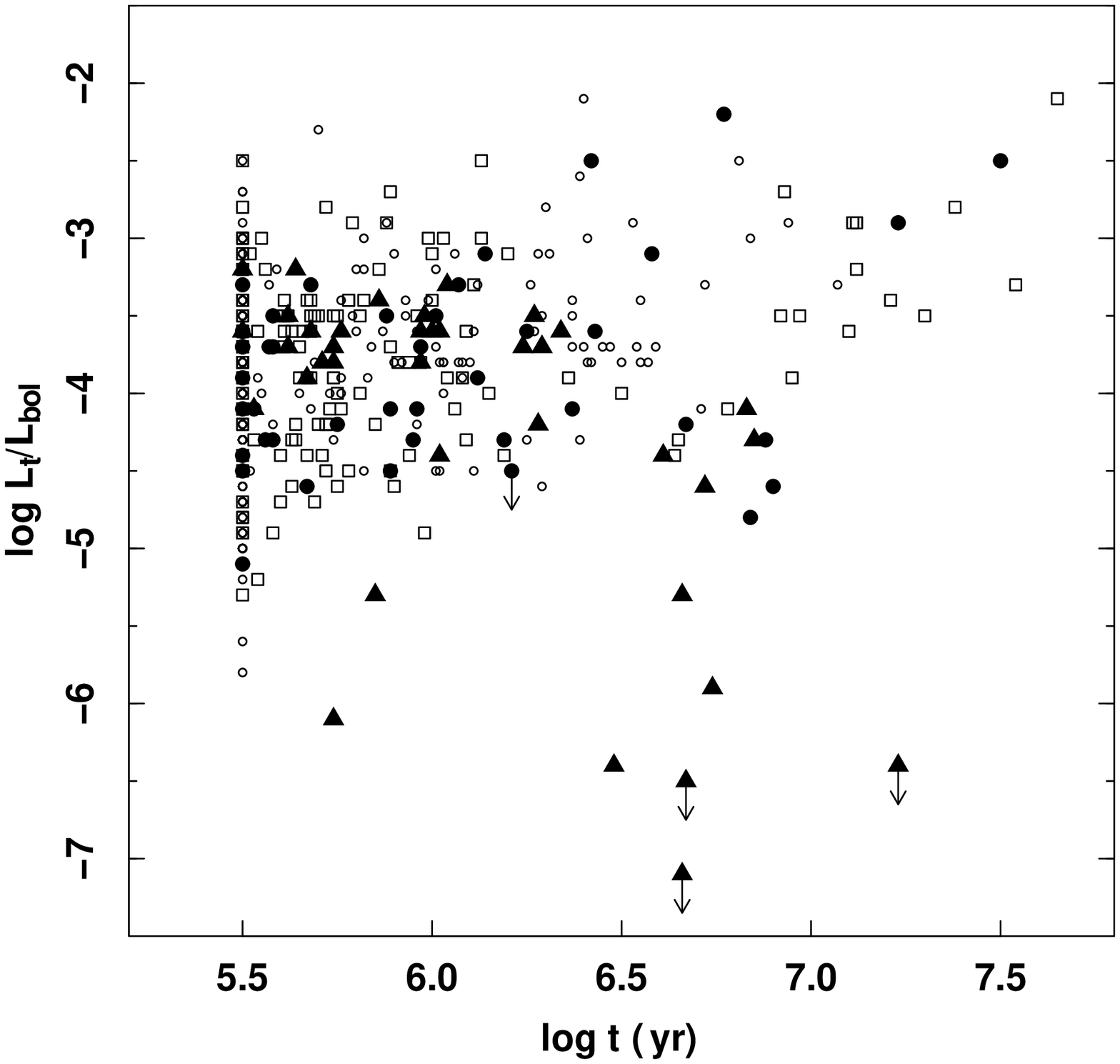}
  \end{minipage} \\ [0.0in]
  \begin{minipage}[t]{1.0\textwidth}
  \centering
  \includegraphics[height=0.45\textheight]{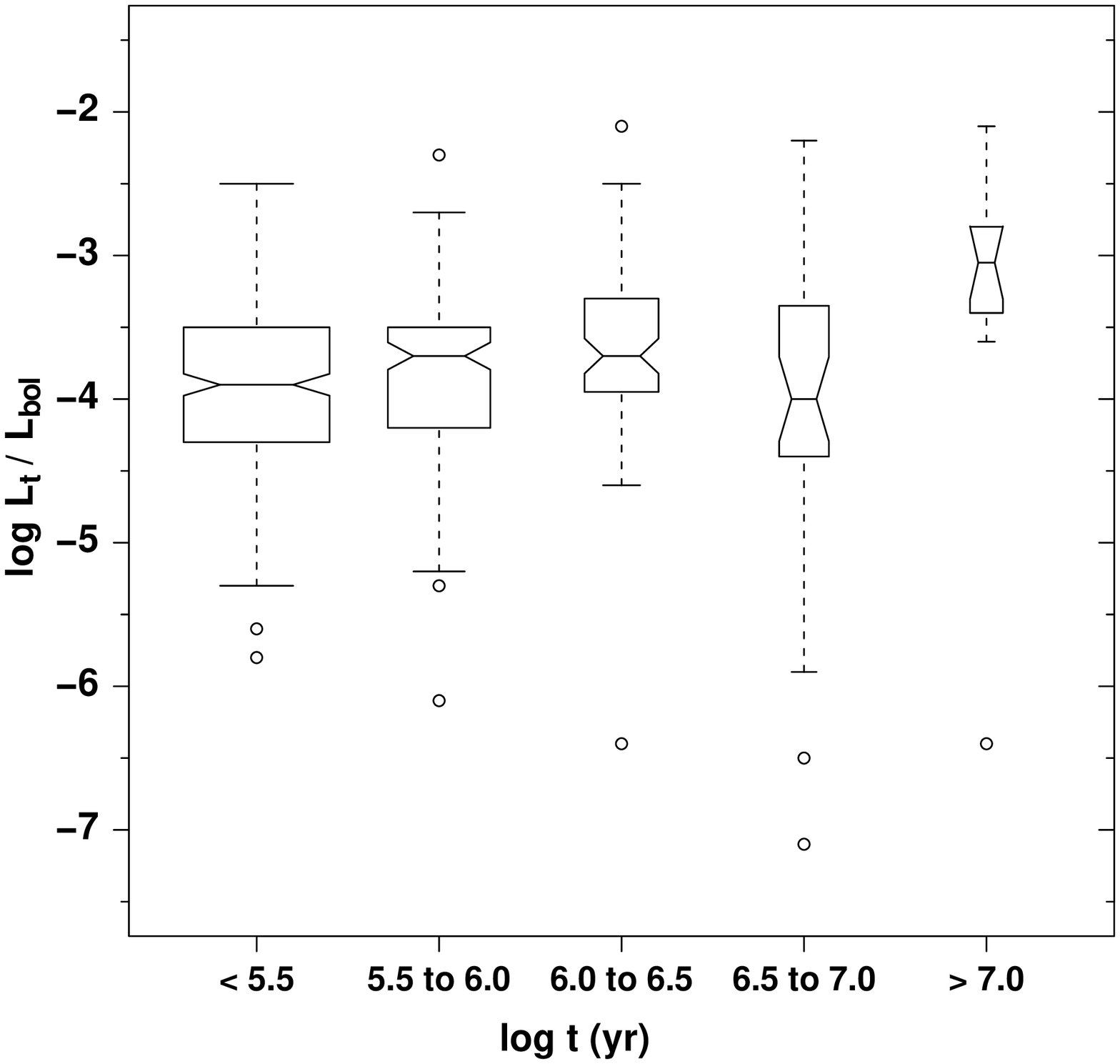}
  \end{minipage}
\end{figure}

\clearpage
\newpage

\begin{figure}
\centering
 \begin{minipage}[t]{1.0\textwidth}
  \centering
  \includegraphics[height=0.45\textheight]{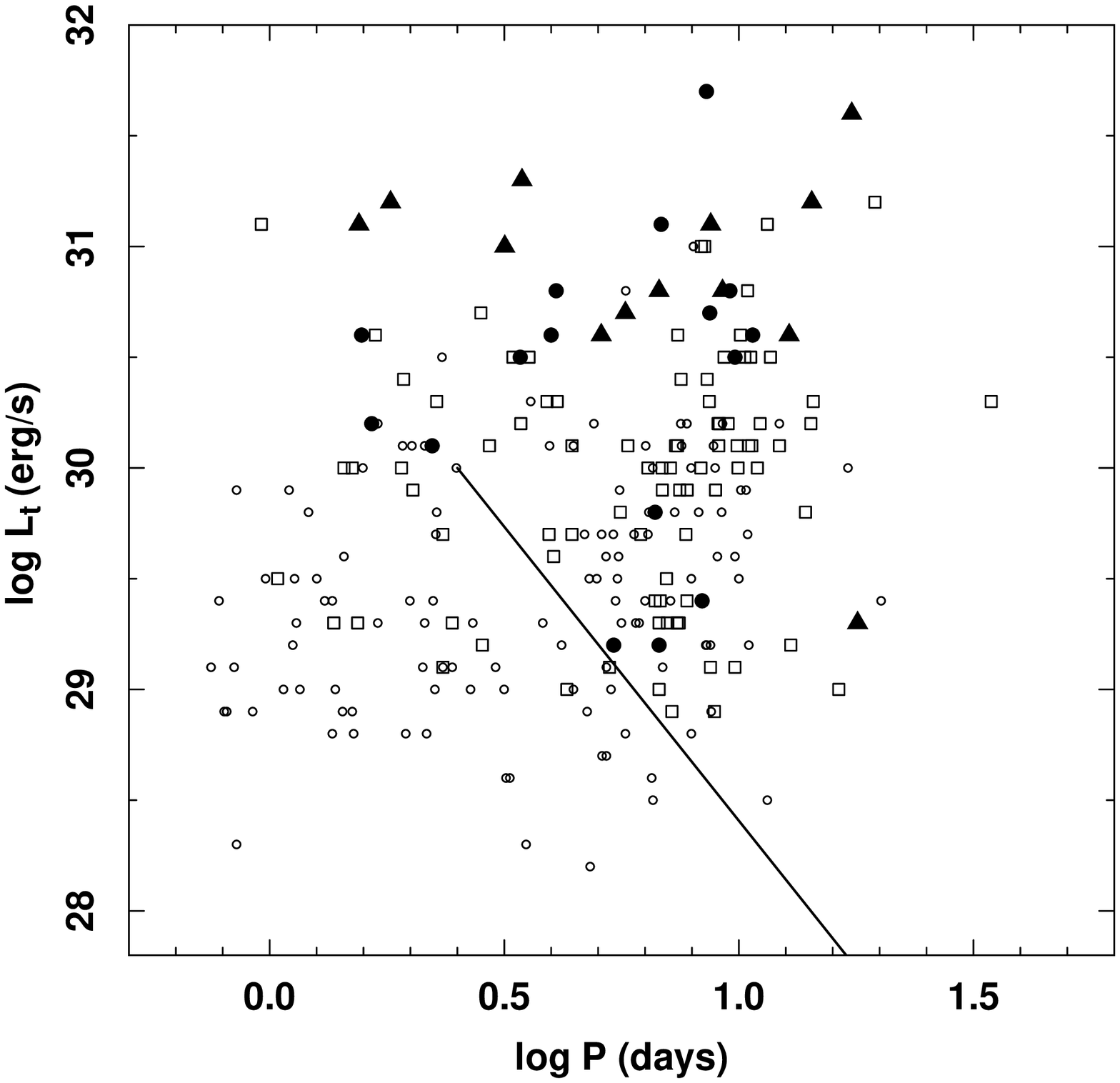}
  \end{minipage} \\ [0.0in]
  \begin{minipage}[t]{1.0\textwidth}
  \centering
  \includegraphics[height=0.45\textheight]{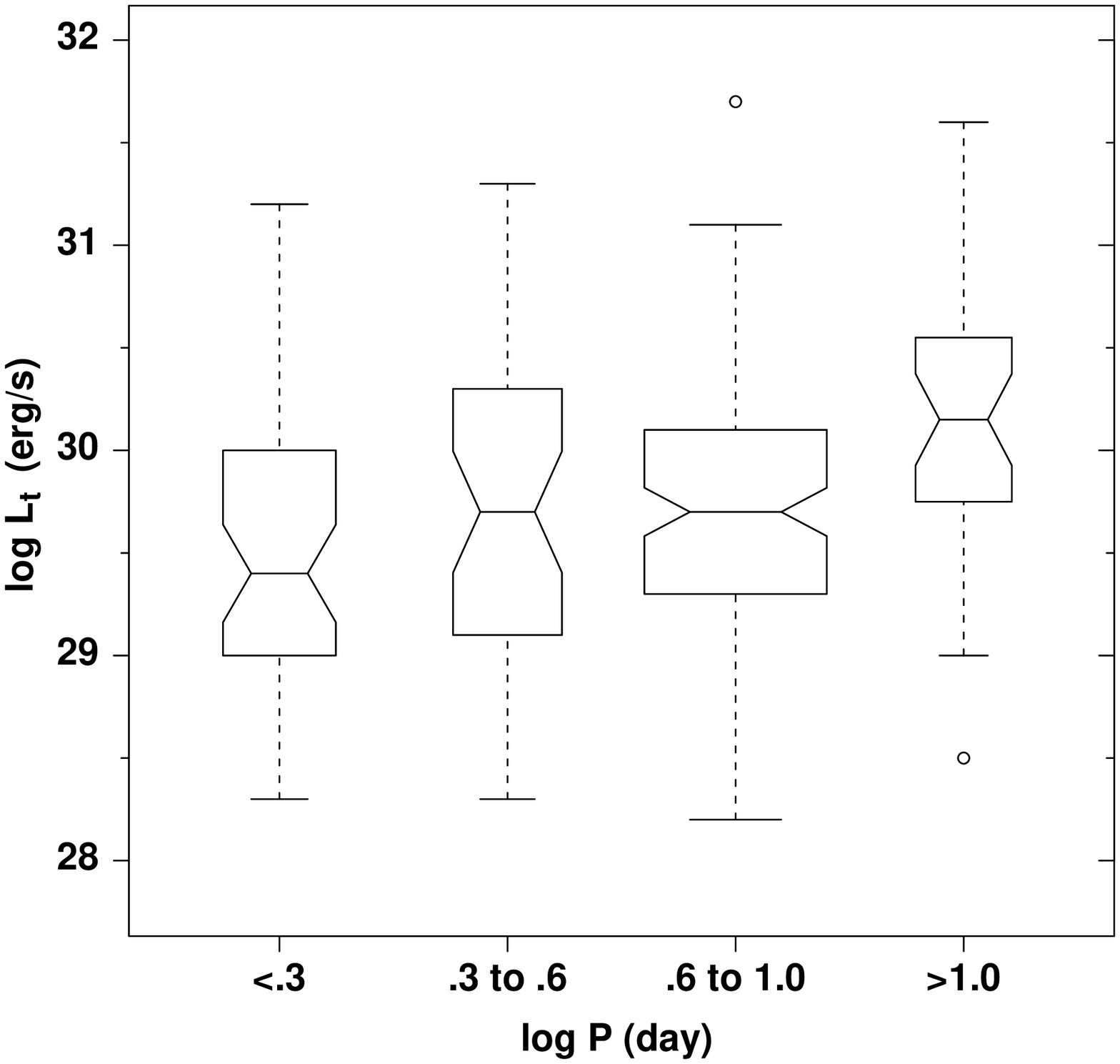}
  \end{minipage}
\caption{Relationship between PMS X-rays and stellar rotation periods:
(a) scatter plot of $\log L_t$ and $\log P$, (b) boxplot of $\log L_t$
and $\log P$, (c) scatter plot of $\log L_t/L_{bol}$ and $\log P$, (d)
boxplot of $\log L_t/L_{bol}$ and $\log t$.  See Figure
\ref{Lx_Lbol.fig} caption for symbol definitions.  The lines in
panels (a) and (c), reproduced from Figure \ref{main_sequence.fig}a, 
show the relationships seen in solar-mass main sequence stars.  
\label{Lx_rot.fig}}
\end{figure}

\clearpage
\newpage

\begin{figure}
\centering
 \begin{minipage}[t]{1.0\textwidth}
  \centering
  \includegraphics[height=0.45\textheight]{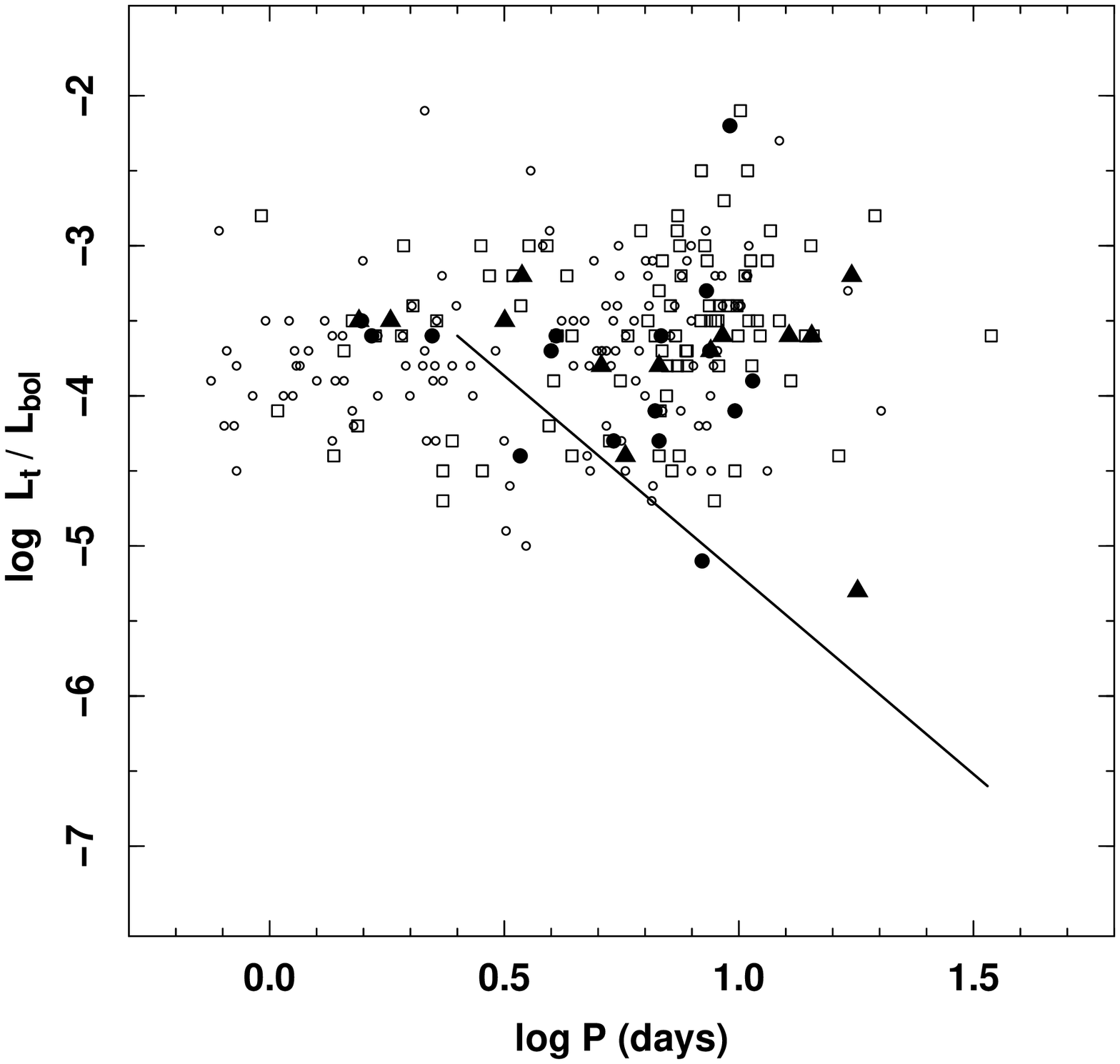}
  \end{minipage} \\ [0.0in]
  \begin{minipage}[t]{1.0\textwidth}
  \centering
  \includegraphics[height=0.45\textheight]{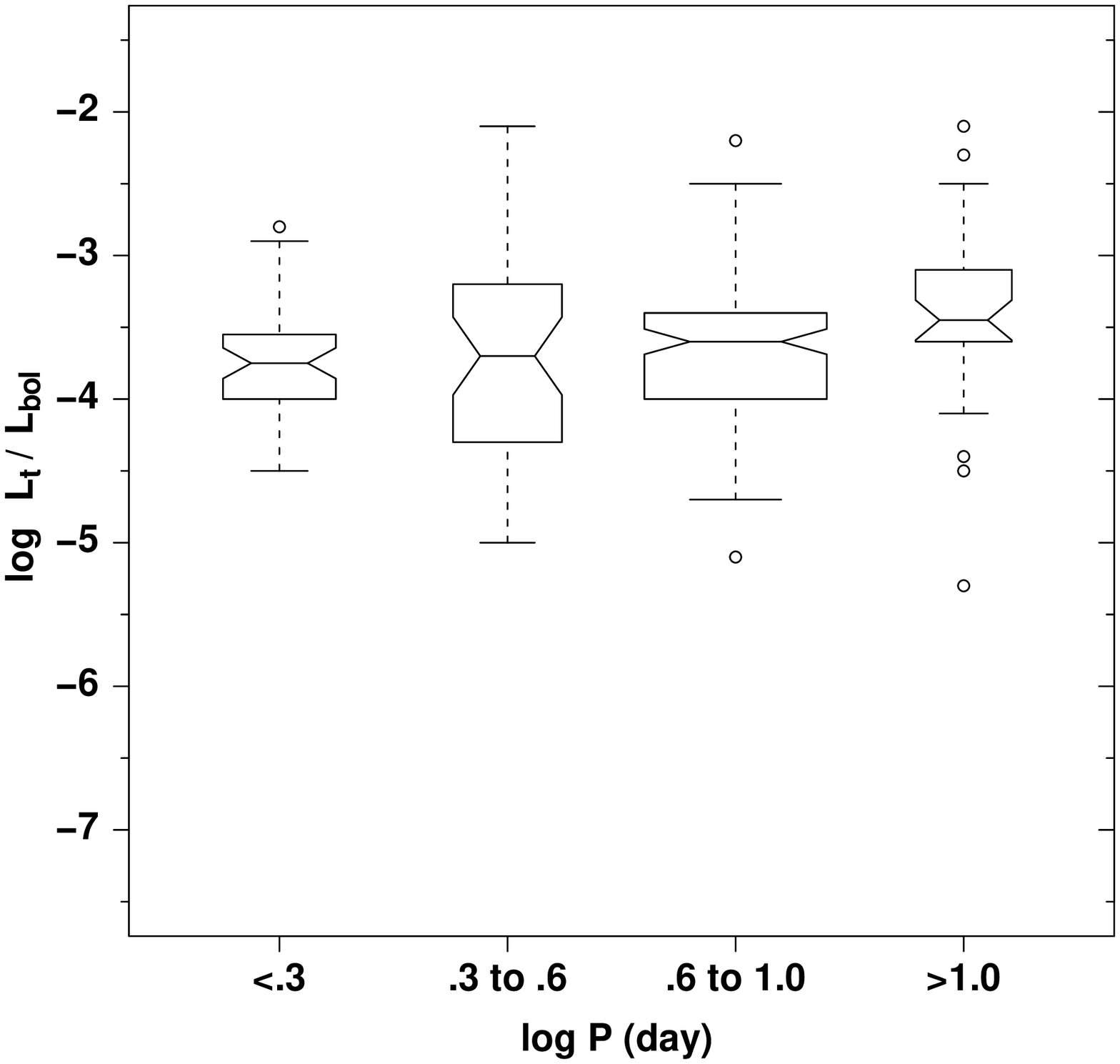}
  \end{minipage}
\end{figure}

\clearpage
\newpage

\begin{figure}
\centering
 \begin{minipage}[t]{1.0\textwidth}
  \centering
  \includegraphics[height=0.45\textheight]{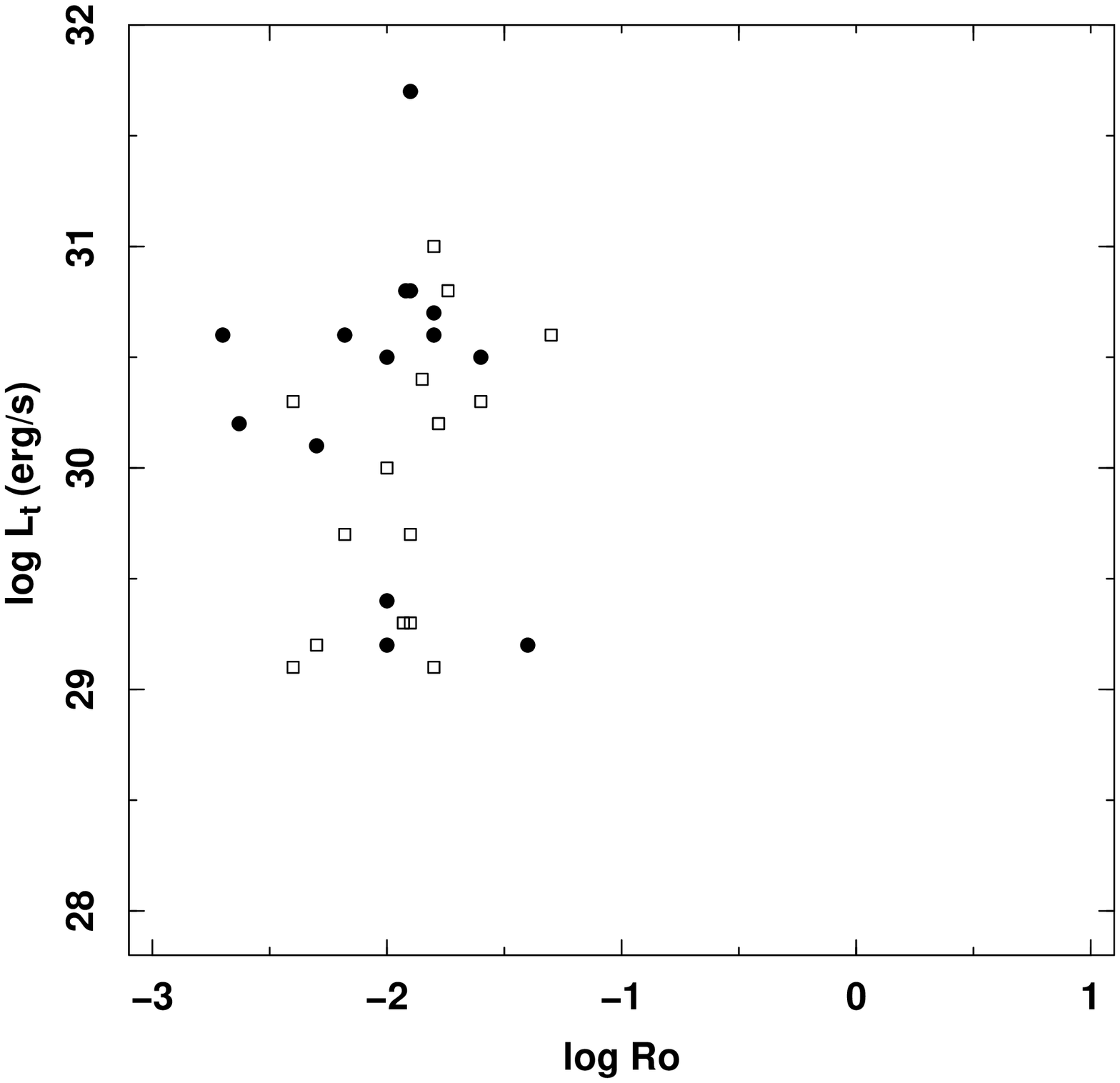}
  \end{minipage} \\ [0.0in]
  \begin{minipage}[t]{1.0\textwidth}
  \centering
  \includegraphics[height=0.45\textheight]{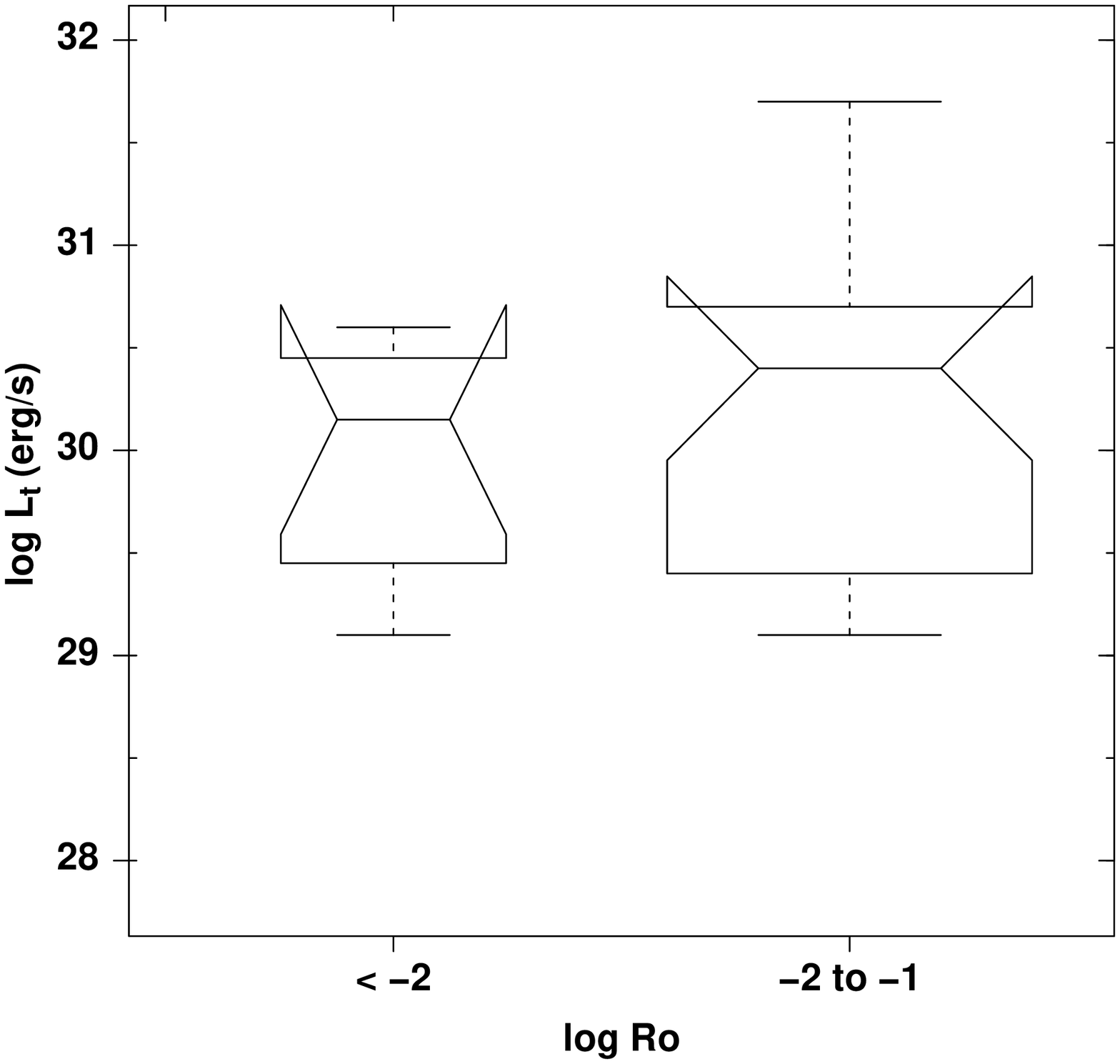}
  \end{minipage}
\caption{Relationship between PMS X-rays and Rossby number: (a) scatter
plot of $\log L_t$ and $\log Ro$; (b) boxplot of $\log L_t$ and $\log
Ro$, (c) scatter plot of $\log L_t/L_{bol}$ and $\log Ro$, (d)
boxplot of $\log L_t/L_{bol}$ and $\log Ro$.  This plot is restricted
to stellar masses $0.5-1.2$ M$_\odot$ for which Rossby numbers have
been calculated \citep{Kim96}. See Figure \ref{Lx_Lbol.fig} caption for
symbol definitions. The lines, reproduced from Figure
\ref{main_sequence.fig}b, show the relationship seen in main sequence
stars.
\label{Lx_Rossby.fig}}
\end{figure}

\clearpage
\newpage

\begin{figure}
\centering
 \begin{minipage}[t]{1.0\textwidth}
  \centering
  \includegraphics[height=0.45\textheight]{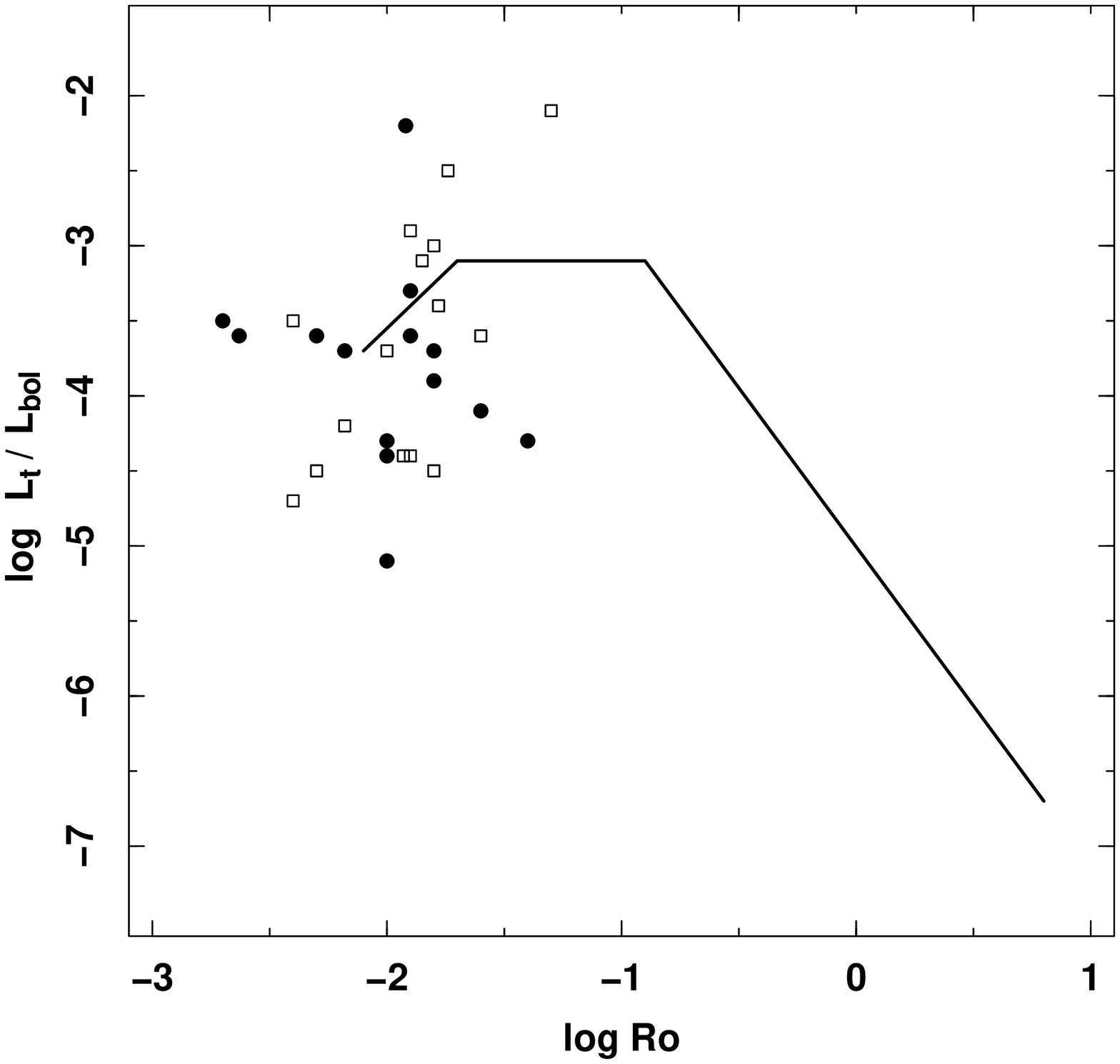}
  \end{minipage} \\ [0.0in]
  \begin{minipage}[t]{1.0\textwidth}
  \centering
  \includegraphics[height=0.45\textheight]{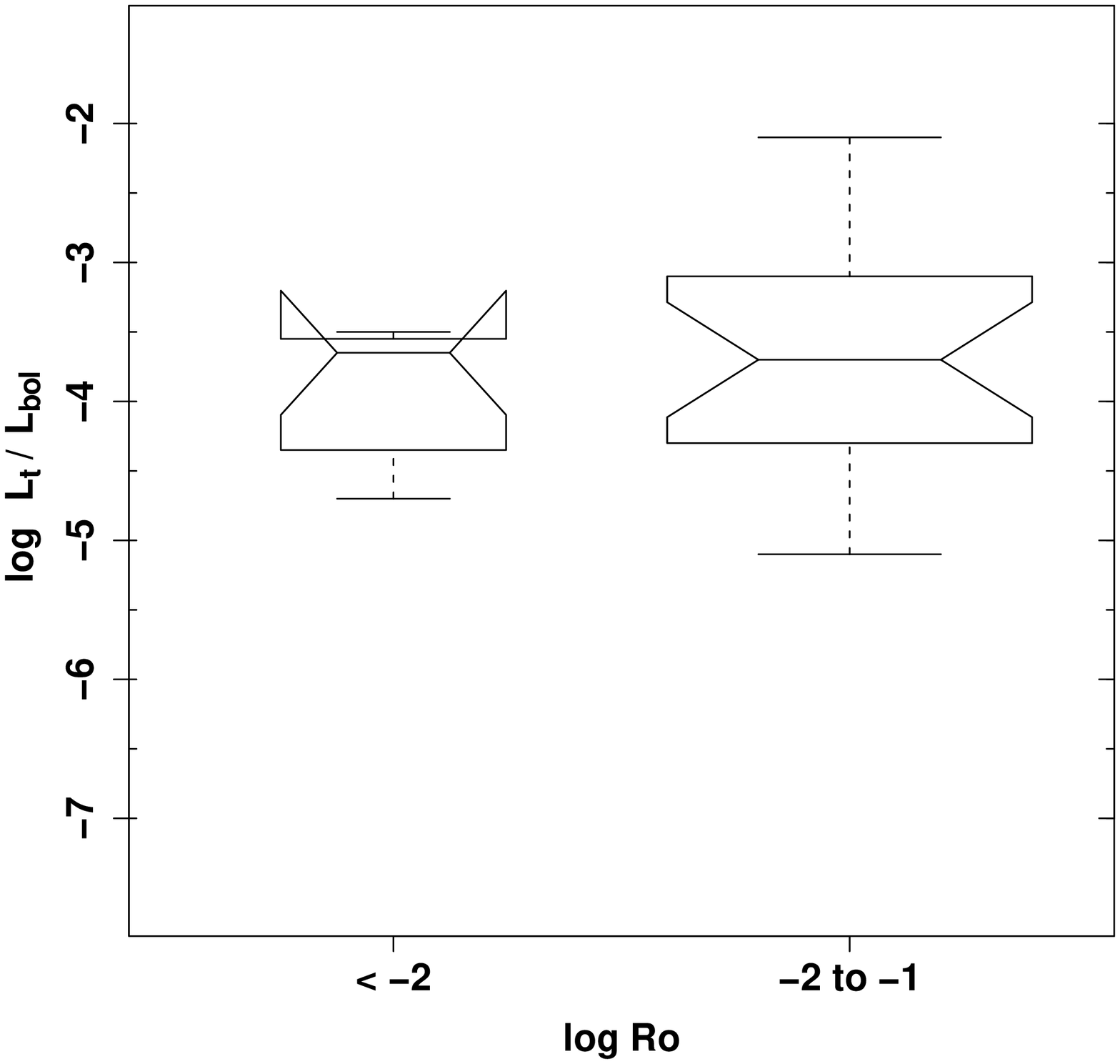}
  \end{minipage}
\end{figure}

\clearpage
\newpage

\begin{deluxetable}{ccrrrrccrccc}
\tabletypesize{\scriptsize}
\tablewidth{0pt}
\tablecolumns{12}
\tablecaption{X-ray properties of well-characterized ONC stars (sample)
\label{srcs.tab}}

\tablehead{
\multicolumn{2}{c}{ACIS source} &
\multicolumn{7}{c}{Stellar properties} &
\multicolumn{3}{c}{X-ray properties} \\

\colhead{CXOONC J} &
\colhead{Star} &
\colhead{$\log L_{bol}$} &
\colhead{$\log M$} &
\colhead{$\log t$} &
\colhead{Disk} &
\colhead{Period} &
\colhead{Period} &
\colhead{$Ro$} &
\colhead{$\log L_s$} &
\colhead{$\log L_t$} &
\colhead{$\log$} \\

\colhead{} &
\colhead{} &
\colhead{($L_\odot$)} &
\colhead{($M_\odot$)} &
\colhead{(yr)} &
\colhead{} &
\colhead{(day)} &
\colhead{Ref.} &
\colhead{} &
\colhead{(erg s$^{-1}$)} &
\colhead{(erg s$^{-1}$)} &
\colhead{$L_t/L_{bol}$} 
}
\startdata
053510.5-052245  &  JW 345  &    0.43  &  -0.66  &  5.50  &    1  &    8.21  &  H     & \nodata  &  29.7  &  29.8  &  -4.2 \\
053510.7-052344  &  JW 352  &    1.19  &  -0.74  &  5.50  &    1  &    8.00  &   H    & \nodata  &  30.8  &  31.0  &  -3.8 \\
053510.7-052628  &  JW 354  &   -0.76  &  -0.77  &  6.03  &    1  &\nodata   & \nodata& \nodata  &  28.8  &  28.8  &  -4.0 \\
053510.8-052759  &  JW 357  &   -0.56  &  -0.89  &  5.50  &   -1  & \nodata  & \nodata&\nodata   &  29.2  &  29.2  &  -3.8 \\
053510.9-052448  &  JW 356  &   -0.43  &  -0.77  &  5.50  &    1  &    4.69  &  H     & \nodata  &  29.5  &  29.7  &  -3.5 \\
053511.2-051720  &  JW 358  &   -0.07  &  -0.57  &  5.50  &    1  &    4.03  &  H     & \nodata  &  29.2  &  29.6  &  -3.9 \\
053511.4-051401  &  H 3005  &   -0.70  &  -0.89  &  5.50  &   -1  &\nodata   & \nodata& \nodata  &  29.5  &  29.9  &  -3.0 \\
053511.4-051911  &  JW 361  &   -0.33  &  -0.49  &  5.86  &    0  &    2.94  &  H     & \nodata  &  29.8  &  30.1  &  -3.2 \\
053511.4-052602  &  JW 365  &    0.85  &   0.09  &  5.50  &    1  &    4.08  &  H     &    -1.9  &  30.5  &  30.8  &  -3.6 \\
053511.6-052421  &  JW 366  &   -0.35  &  -0.92  &  5.50  &   -1  &\nodata   & \nodata& \nodata  &  28.0  &  28.0  &  -5.2 \\
\enddata

\tablecomments{The full table is available only on-line as a machine-readable table}

\end{deluxetable}

\clearpage

\end{document}